\documentclass{article}
\usepackage{spconf,amsmath,graphicx}
\usepackage{lettrine}
\usepackage{color}
\usepackage{multirow}
\usepackage{booktabs}
\usepackage{subfigure}
\graphicspath{{figures/}}

\renewcommand{\paragraph}[1]{{\vspace{+2mm}\noindent\bf #1}}

\title{Multispectral Focal Stack Acquisition Using A Chromatic Aberration Enlarged Camera}
\name{Qian Huang{$^1$}, Yunqian Li{$^1$}, Linsen Chen{$^1$}\thanks{We would like to acknowledge funding from NSFC Projects 61671236, 61631009, 61571215, 61422107, 61371166 and 61627804, National Science Foundation for Young Scholar of Jiangsu Province, China (Grant No. BK20160634 and No. BK20140610), along with the National key foundation
for exploring scientific instrument No.2013YQ140517.}, Xiaoming Zhong{$^2$}, Jinli Suo{$^3$}, Zhan Ma{$^1$}, Tao Yue{$^1$}, Xun Cao{$^1$}}
\address{{$^1$}School of Electronic Science and Technology, Nanjing University, China\\
  {$^2$}Beijing Institute of Space Mechanics and Electricity, China\\
{$^3$}Department of Automation, Tsinghua University, China}
\begin{document}
%
\maketitle
\begin{abstract}
Capturing more information, e.g. geometry and material, using optical cameras can greatly help the perception and understanding of complex scenes. This paper proposes a novel method to capture the spectral and light field information simultaneously. By using a delicately designed chromatic aberration enlarged camera, the spectral-varying slices at different depths of the scene can be easily captured. Afterwards, the multispectral focal stack, which is composed of a stack of multispectral slice images focusing on different depths, can be recovered from the spectral-varying slices by using a Local Linear Transformation (LLT) based algorithm. The experiments verify the effectiveness of the proposed method. 
\end{abstract}
\section{Introduction}
\label{sec:intro}

\lettrine[lines=2]{T}{o} perceive and understand more complex scenes, researchers attempt to capture more information, e.g. multispectral and light field, of scenes. Traditional cameras take pictures by using 2D sensors with Bayer filter arrays. The captured trichromatic images are 2D projections of 3D scenes with three color channels, i.e. red, green and blue. With the development of optics and computational photography, both multispectral and light field acquisition have been widely explored to capture more spectral channels or light field information of scenes. 

In the past decades, several spectral imaging methods were proposed to capture color images with more spectral channels than traditional trichromatic photography. Generally, according to the system architectures, existing multispectral cameras can be divided into several types, e.g., scanning based spectrometers \cite{james2007spectrograph}, filter-based spectrometers\cite{gat2000imaging}, Coded Aperture Snapshot Spectral Imager(CASSI) \cite{arce2014compressive}, Computed Tomography Imaging Spectrometer(CTIS) \cite{vandervlugt2007reconfigurable}, and Prism-Mask Video Imaging Spectrometer(PMVIS) \cite{cao2011prism}.

Besides spectral information, the light field (or equivalently, scene depth) is also an important clue for many tasks in computer vision and graphics. 
Recently, to capture the light field, several methods, e.g., microlens array based method \cite{ng2005light}, multi-camera array based methods \cite{zhang2004self} and focal stack based methods \cite{li2014saliency} are proposed. (Note that the focal stack is one type of representations of the light field, thus in this paper we use the terms light field and focal stack indiscriminatively.)  Confocal laser scanning microscope(CLSM) \cite{Pawley1995Handbook} can acquire the microscopic focal stack using the scanning scheme. As for snapshot depth acquisition, the chromatic information is explored for extracting depth from color (RGB) images \cite{garcia2000chromatic}\cite{levin2007image}\cite{cossairt2010spectral}\cite{trouve2013passive}. Besides, Time of Flight (ToF) camera \cite{Schuon2008High} and coded illumination \cite{Schubert1997Fast} camera were presented to capture the depth directly, so that the light field can be derived by model based rendering.

\begin{figure}[!ht]
  \vspace{-3mm}
  \begin{minipage}[b]{1.0\linewidth}
  \centerline{\includegraphics[width=7cm]{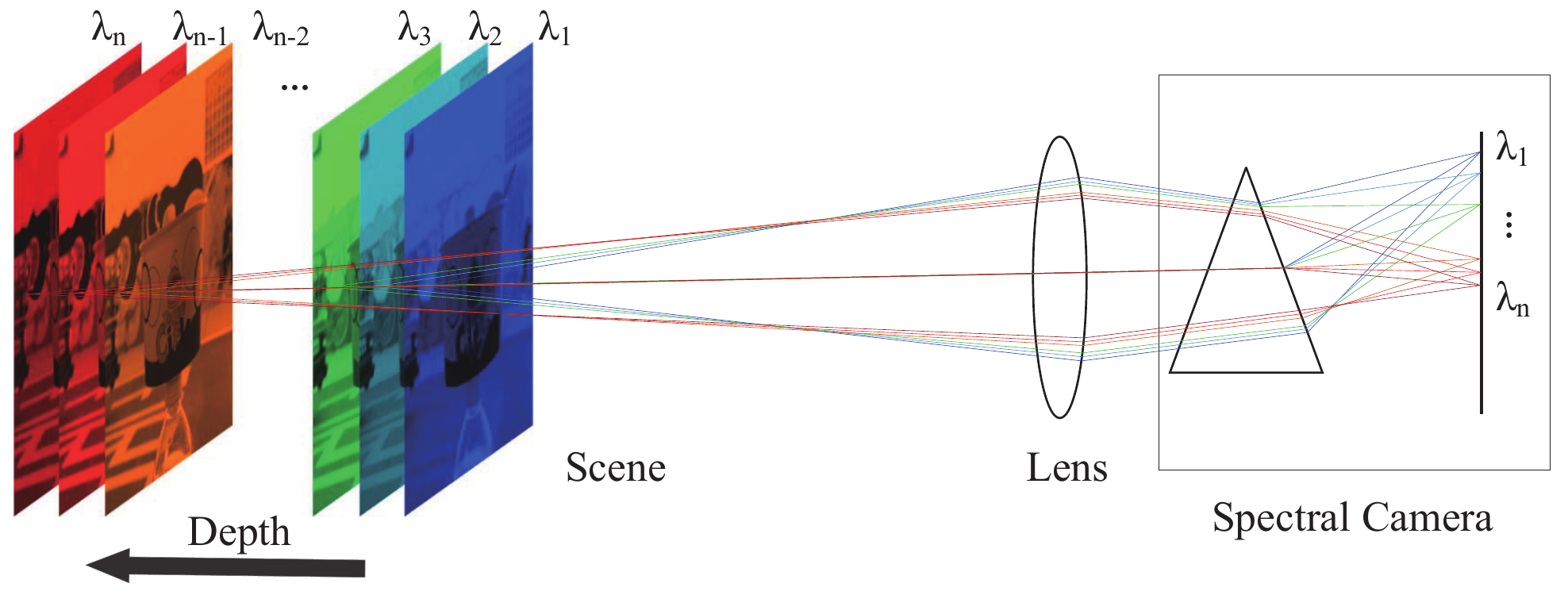}}

\end{minipage}
\caption{Diagram of the proposed multispectral light field acquisition method.}
\label{fig: thumbnail}
  \vspace{-3mm}
\end{figure}

However, capturing both spectral and light field information together is difficult for too many anisotropic data need to be measured. 
To solve this problem, this paper proposes a delicately designed camera system, which tries to enlarge the chromatic aberration of the lens while eliminating its rest aberrations (e.g., spherical aberration, coma aberration and astigmatism). By using the proposed camera, the light rays with different wavelengths from slices at different depths of scenes can focus on the same imaging plane. Thus by dispersing the light incidented into the sensor plane into different spectral channels, the images focusing at different depths can be separated. Placing a multispectral imager at the sensor plane, we can capture a multispectral image whose channels are focused at different depths. In other words, we drive a spectral-varying focal stack whose slices are of different spectral channels. Then, a Local Linear Transformation (LLT) based algorithm is present to reconstruct the multispectral focal stack, which contains full information of multispectral light field. Our optical system model can be simplified in Fig.~\ref{fig: thumbnail}. 

After acquiring the spectrally varying focal stack, we proposed to transfer spectral information between different slices and fill up the vacant channels of each slice to reconstruct the multispectral focal stack. In this paper, inspired by Local Linear Transform (LLT) method introduced by Yue \textit{et al}\cite{yue2014deblur}, we develop a new method called the local linear transformation (LLT) to transfer the spectral information between channels. Specifically, we impose a strong global blur(e.g. Gaussian) on all the captured channels to remove the different high frequency between these slices caused by different focusing planes. There exists a reasonable linear mapping, namely Local Linear Transformation (LLT), between any two blurred channels and the transformation is also valid for the corresponding sharp slices. In order to extract the LLT mappings, the gradient descent based algorithm is applied.

In all, the main contributions of this paper are as follows: (a) a simple chromatic aberration enlarged camera design for multispectral light field acquisition; (b) a local linear transformation(LLT) based reconstruction method for effectively and efficiently reconstructing the multispectral light field.

\section{Proposed Method}

\subsection{Optical System Design}
\begin{figure}[!ht]
  \vspace{-3mm}
\centering
\includegraphics[width=0.9\linewidth]{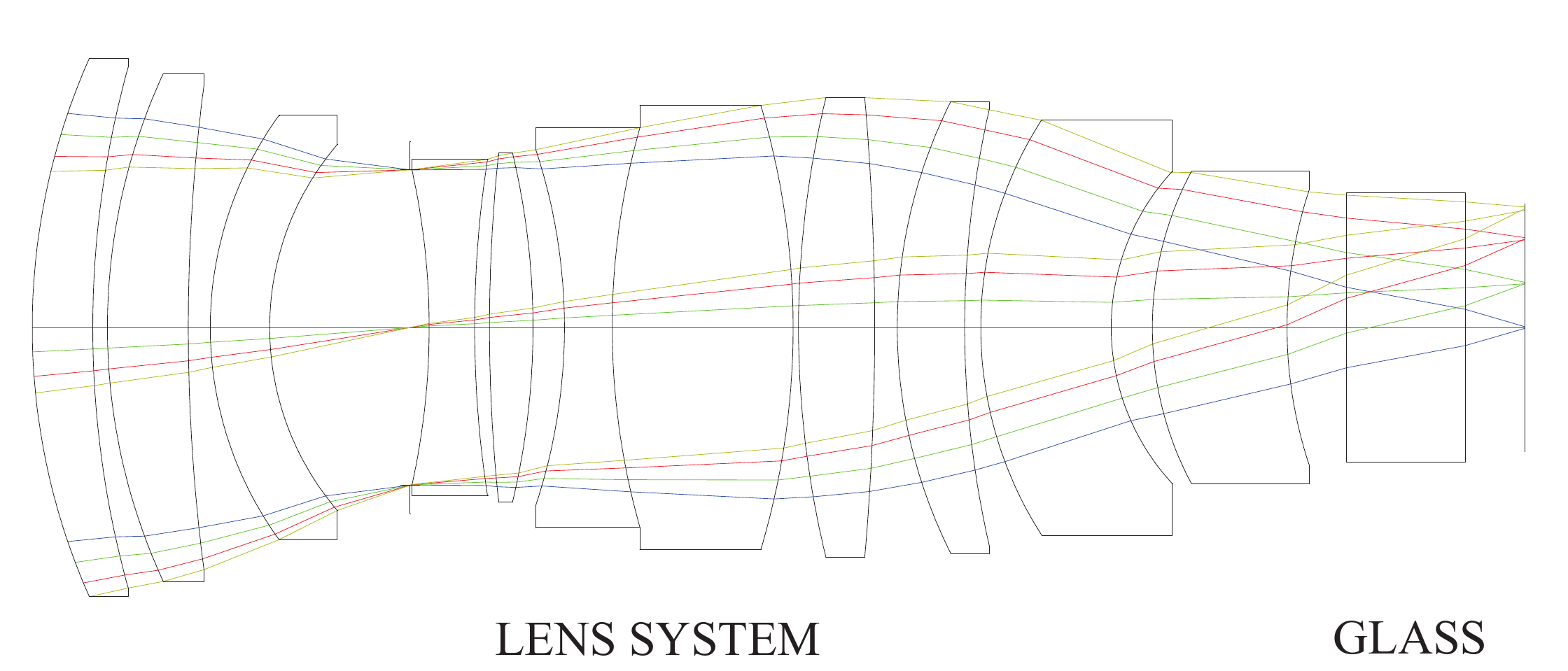}
\caption{\bf The specially designed lens set. (Color rays by: Fields)}
\label{fig: lens}
  \vspace{-3mm}
\end{figure}

\begin{figure}[!ht]
  \vspace{-3mm}
\centering
\includegraphics[width=0.9\linewidth]{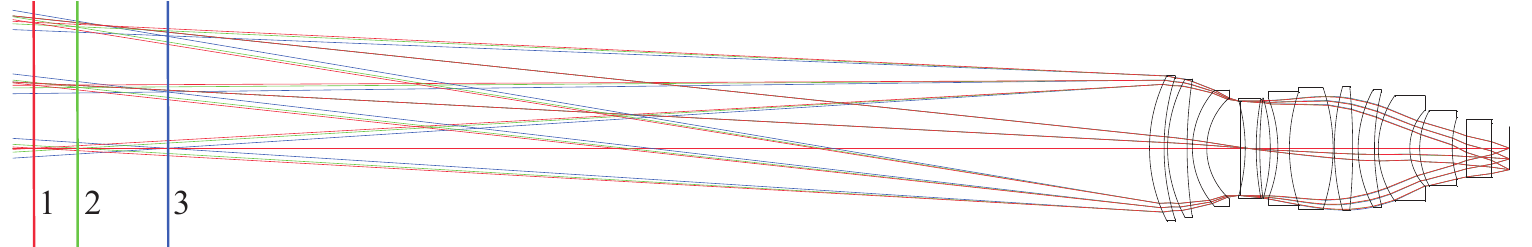}
\caption{\textbf{Different focal planes at different depths in the scene. (Color rays by: Wavelengths)}
Object planes marked 1, 2, 3 denote depths of $d_1$, $d_2$, $d_3$ and wavelengths of 400nm, 550nm, 700nm respectively.}
\label{fig: focal stack}
  \vspace{-3mm}
\end{figure}

In this paper, we propose to design an optical system to focus on planes of different depths of the scene with different spectral channels, so that we can capture the spectrally varying focal stack by using a multispectral imager at once. According to this idea, we want to enlarge the camera's chromatic aberration to enlarge the focusing range while eliminating the rest optical aberrations, including piston, tilt, defocus, spherical aberration, coma, astigmatism, field curvature and image distortion, to get promising image quality of all the channels. Here, instead of starting from the scratch to design the lens, we propose a more simple method, i.e. adding a cubic glass behind a well-designed lens (all the aberrations are well corrected), as shown in Fig.~\ref{fig: lens}. With this specially designed lens set, the imaging system has different focal planes with different depths in the scene. The simulation results with ZEMAX is fundamentally consistent with our theory, as shown in Fig.~\ref{fig: focal stack}. By selecting the central wavelengths of spectral channels ranging from 430nm to 700nm, the corresponding slices between plane 1 and plane 3 in Fig.~\ref{fig: focal stack} can be captured. In our paper, ten spectral channels are selected with equal intervals between the central wavelengths, so that the corresponding slices have non-uniform depth intervals. 

\begin{figure}[!ht]
  \vspace{-3mm}
\centering
\subfigure[]
{
\includegraphics[width=0.32\linewidth]{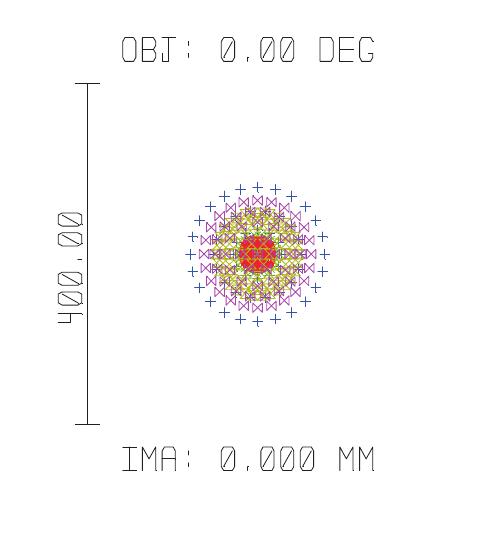}
}
\hfill
\subfigure[]
{
\includegraphics[width=0.58\linewidth]{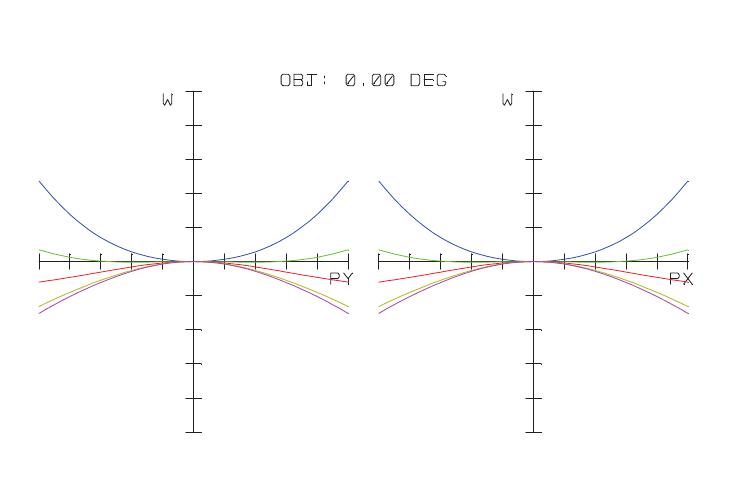}
}
\caption{{\bf Diagram of optical aberrations.} (a)Section of the Spot diagram (b)Section of the OPD.}
\label{fig: aberration}
  \vspace{-3mm}
\end{figure}

Fig.~\ref{fig: aberration}(a) presents the spots pattern of a certain point with different wavelengths. In the shown section of spot diagram, the red rays focus nearly on the best image plane because its RMS(Root Mean Square) radius is the smallest, while other wavelengths from the same depth in the scene are not. This is a result of chromatic aberration. More specifically, the longer the wavelength is, the smaller the RMS radius of the will be, which is especially obvious for the blue and red one in the figure. Fig.~\ref{fig: aberration}(b) represents the optical path difference(OPD) of the system. Taking the object at the spindle for instance still, the disparity(i.e. the OPD of different wavelengths) becomes apparent among different wavelengths as the entrance pupil scalar(i.e. abscissa) increases, which shows the enhanced chromatic aberration indirectly. In other words, when the light deviates the spindle of lens set, the real convergence points of different spectra become more distant spatially. Note that different colors in Fig.~\ref{fig: aberration} represent different wavelengths varying from 430nm to 700nm.


\subsection{Multispectral Focal Stack Reconstruction}

\begin{figure}[!ht]
\vspace{-3mm}
\begin{minipage}[b]{1.0\linewidth}
  \centerline{\includegraphics[width=8.5cm]{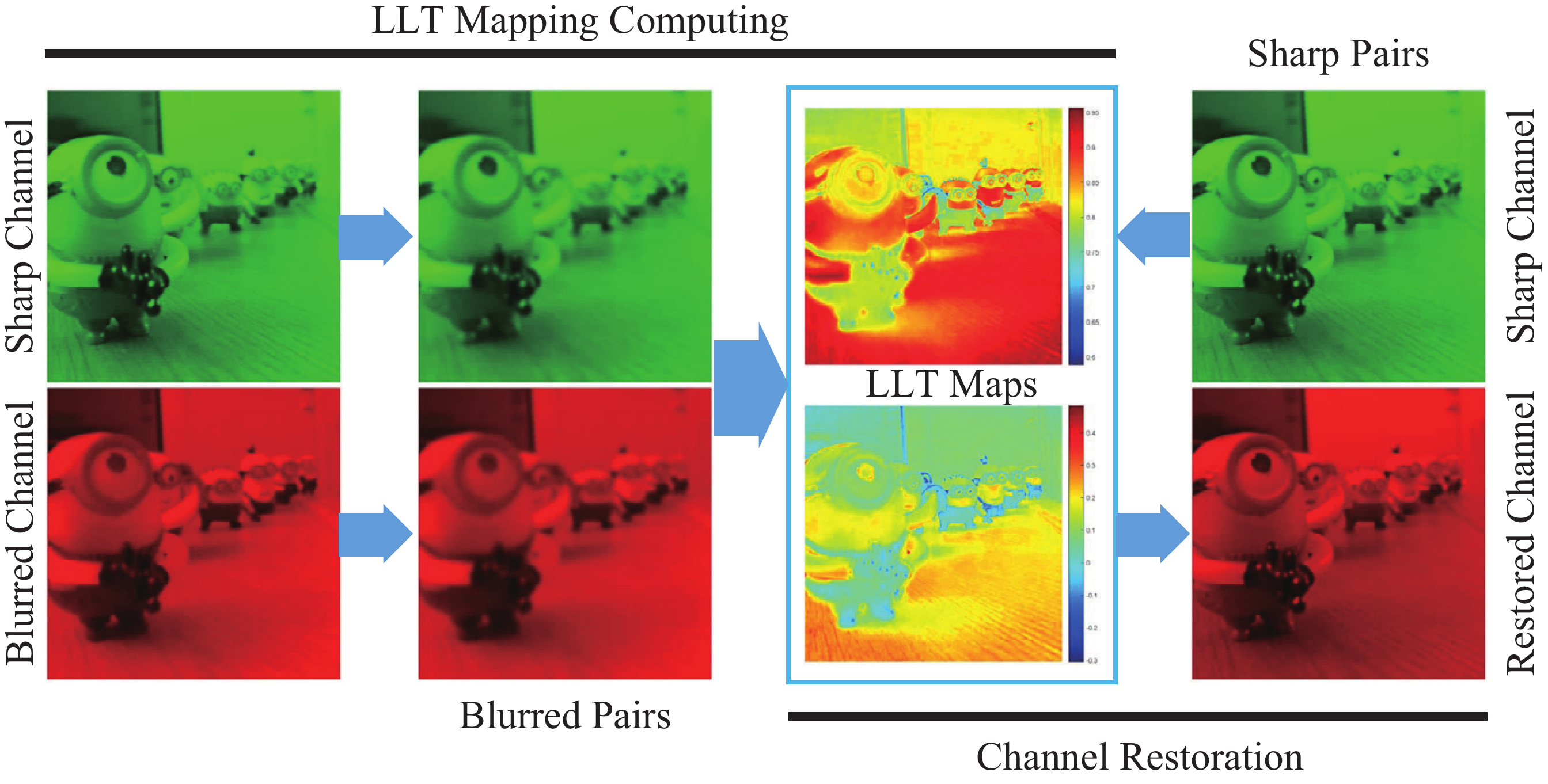}}
\end{minipage}
\caption{Flowchart of our multispectral focal stack reconstruction.}
\label{fig: flowchart}
\vspace{-2mm}
\end{figure}

Here, we present the proposed Local Linear Mapping (LLT) based multispectral focal stack reconstruction algorithm. As shown in Fig.~\ref{fig: flowchart}, we capture one channel at a single depth, which is treated as the sharp channel in LLT algorithm. The blurred channel is derived by blurring the captured image computationally. By computing the LLT maps $\textbf{A}_{i, k}$ and $\textbf{B}_{i, k}$, we can restore the missing channels by channel transferring.

According to Local Linear Transformation (LLT) property introduced by Yue \textit{et al}.\cite{yue2014deblur}, in a local area with the same blur effect, the pixel values of different channels follow a certain linear transformation, which is valid in the same area for pixels of the sharp version of those channels as well. In our scenario, the defocus blur varies in both spatial and channel dimensions. It is no trivial to apply the LLT in this case directly, since there is no obvious blur-sharp pair in our application. Therefore, the LLT is proposed to transfer the spectral information while keeping the blur patterns which imply depth information of the scene.

For any two slices, which are captured with different wavelengths and different local blur effects caused by different focusing distances, we apply a strong \textit{Gaussian Kernel} to blur both of them to remove the different high frequency information, so that the blurred images can be regarded as uniformly blurred with a large Gaussian kernel. Thus, the two slices are of the same blur pattern and the LLT maps can be computed from these blurred slice pairs. Given the LLT Maps, the full-channel focal stacks can be restored by linearly transforming between different channels of the captured spectral-varying focal stack.

Specifically, we blur all the channels with a large Gaussian Kernel (${\bf G}_{\sigma}$) at first. Empirically, the standard variation $\sigma = 10$ is strong enough for images captured in practice. Then, according to LLT property \cite{yue2014deblur}, we know there exist local linear transforms between blurred channels and the original channels. The relationship can be described as follows:
\begin{subequations}
\begin{equation}
\begin{aligned}
\textbf{I}_{\lambda_i} = \textbf{A}_{i, k}\odot \textbf{I}_{\lambda_k} + \textbf{B}_{i, k}
\end{aligned}
\end{equation}
\begin{equation}
\begin{aligned}
\textbf{L}_{d_k, \lambda_i} = \textbf{A}_{i, k}\odot \textbf{L}_{d_k, \lambda_k} + \textbf{B}_{i, k},
\end{aligned}
\label{restoreEquation}
\end{equation}
\end{subequations}
where $\textbf{I}_{\lambda_i}$ and $\textbf{I}_{\lambda_k}$ are two blurred channels, which are computed by blurring the original sharp slice pair $\textbf{L}_{d_i, \lambda_i}$ and $\textbf{L}_{d_k, \lambda_k}$. Since the Gaussian blur kernel applied here are very large, the blurred images can be regarded as uniformly blurred. Thus $\textbf{I}_{\lambda_i}$ and $\textbf{I}_{\lambda_k}$ do not contain the depth information, so we remove the subscript $d_i$ and $d_k$ here.  $\textbf{L}_{d_k, \lambda_i}$ and $\textbf{L}_{d_k, \lambda_k}$ are the corresponding sharp channels. $\textbf{A}_{i, k}$, $\textbf{B}_{i, k}$ are the local linear transformation maps. $d_k$ and $\lambda_i$ represent the depth in the scene and corresponding wavelength separately. $\odot$ means element-wise multiplication of matrices.

To compute LLT maps $\textbf{A}_{i, k}$ and $\textbf{B}_{i, k}$, an objective function is introduced:
\begin{equation}
\label{ObFunction}
\begin{aligned}
\mathop{\min} E &=\|\textbf{A}_{i, k}\odot\textbf{I}_{\lambda_k}+\textbf{B}_{i, k}-\textbf{I}_{\lambda_i}\|^2_2\\
&+\alpha\|\textbf{A}_{i, k}\odot\nabla\textbf{I}_{\lambda_k}-\nabla\textbf{I}_{\lambda_i}\|^2_2\\
&+\beta(\|\nabla\textbf{A}_{i, k}\|^2_2+\|\nabla\textbf{B}_{i, k}\|^2_2),
\end{aligned}
\end{equation}
where $\textbf{I}_S$ and $\textbf{I}_B$ separately represents the original sharp and blurred channel.$\alpha$ and $\beta$ are the weights of constraint terms, and are set to 1 and 0.1 according to \cite{yue2014deblur}. $\nabla$ is the gradient operator. 

The traditional Gradient Descent method is applied to optimize Eq.~\ref{ObFunction}. Specifically, the derivatives of Eq.~\ref{ObFunction} can be computed respectively by
\begin{equation}
\label{derivative}
\begin{aligned}
g_A &= 2 \textbf{I}_S\odot(\textbf{A}_{i, k}\odot\textbf{I}_{\lambda_k}+\textbf{B}_{i, k}-\textbf{I}_{\lambda_i})\\
&+2\alpha\nabla\textbf{I}_S\odot(\textbf{A}_{i, k}\odot\nabla\textbf{I}_{\lambda_k}-\nabla\textbf{I}_{\lambda_i}) + 2\beta\nabla^T\nabla\textbf{A}_{i, k}\\
g_B &= 2 (\textbf{A}_{i, k}\odot\textbf{I}_{\lambda_k}+\textbf{B}_{i, k}-\textbf{I}_{\lambda_i})+2\beta\nabla^T\nabla\textbf{B}_{i, k}, 
\end{aligned}
\end{equation}

\begin{figure}[t]
\vspace{-3mm}
\centering
\subfigure
{
  \includegraphics[width=0.35\linewidth]{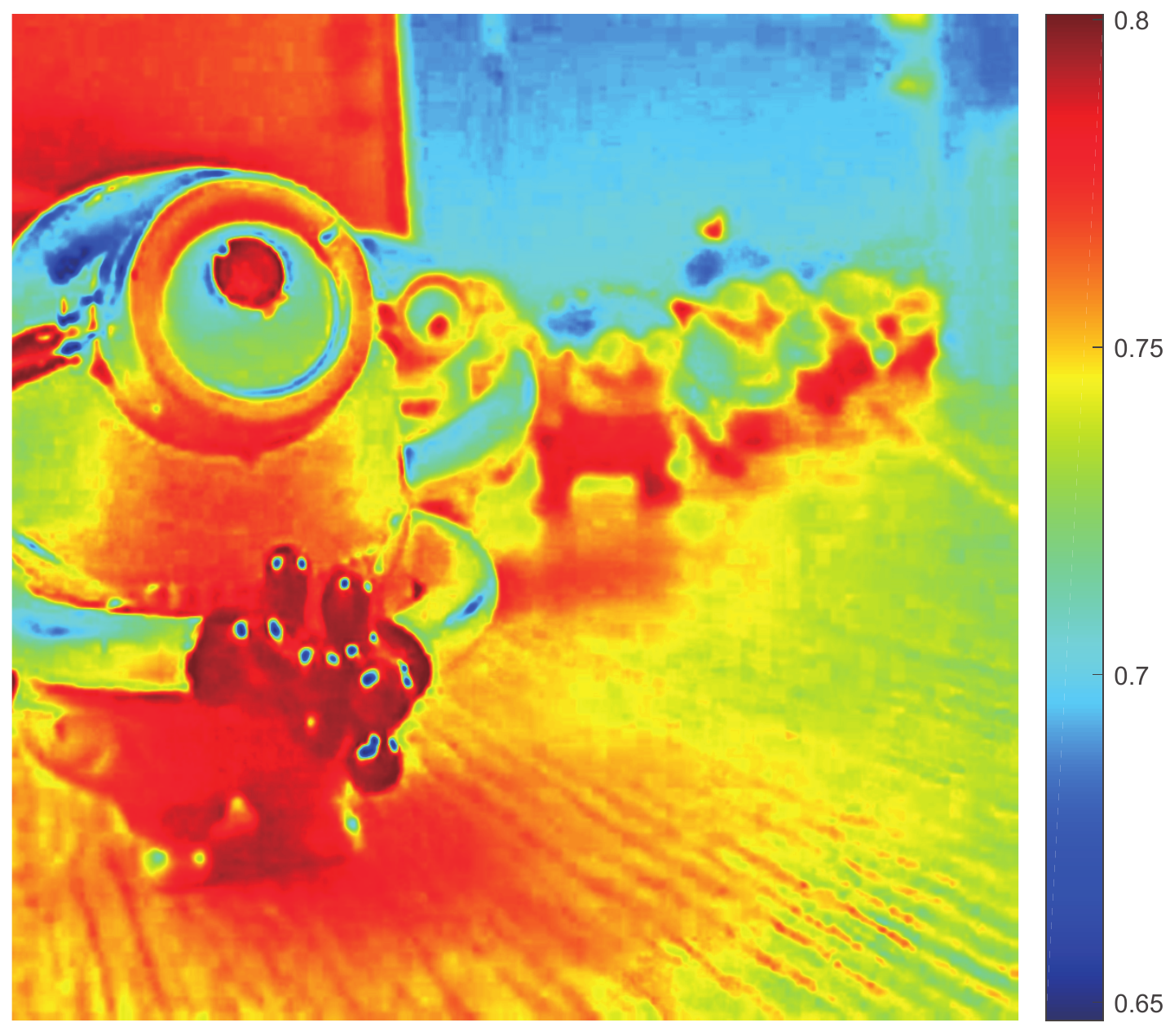}
}
\subfigure
{
  \includegraphics[width=0.35\linewidth]{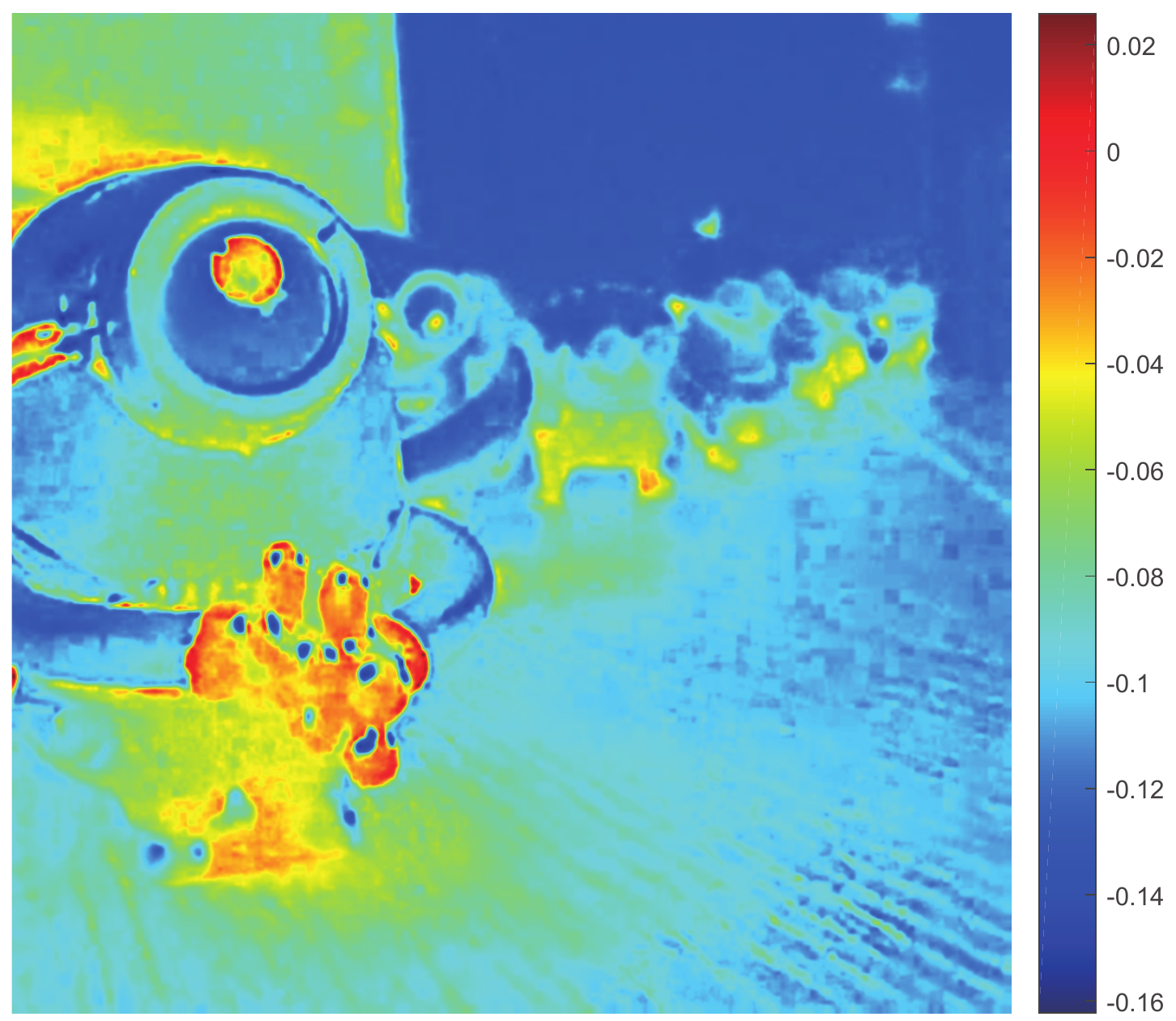}
}

\caption{An example of LLT map pair $\textbf{A}_{i, k}$(left) and $\textbf{B}_{i, k}$(right)}
\label{fig: glt}
\vspace{-2mm}
\end{figure}

By iteratively searching along the gradient directions given by Eq.~\ref{derivative}, $\textbf{A}_{i, k}$ and $\textbf{B}_{i, k}$ can be derived. Fig.~\ref{fig: glt} shows an example of LLT map pair $\textbf{A}_{i, k}$ and $\textbf{B}_{i, k}$.

\begin{figure*}[t]
  \vspace{-3mm}
	\centering
	\subfigure[$d_1$]
	{
		\includegraphics[width=0.16\linewidth]{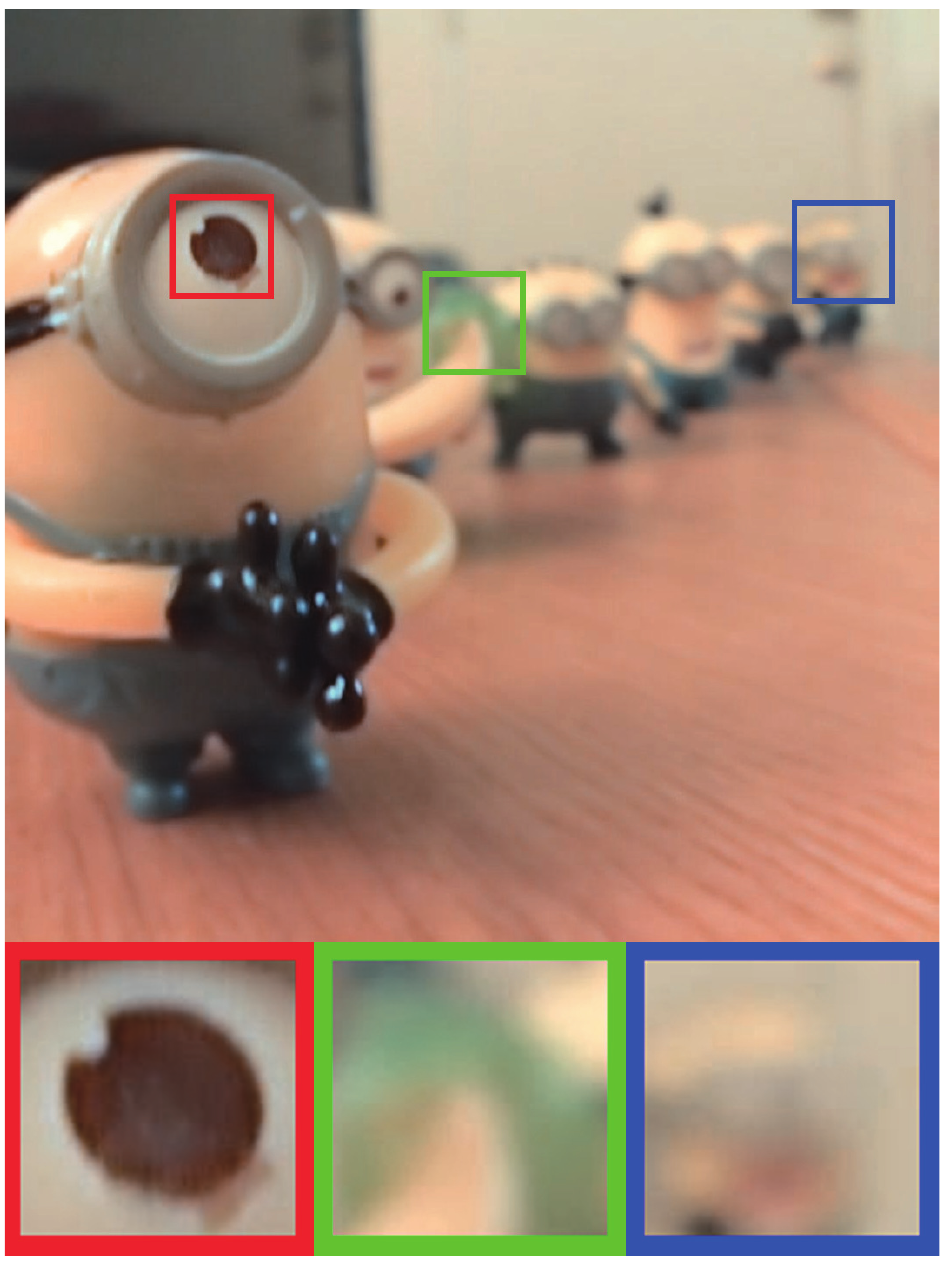}
	}
		\hspace{-2.5ex}
	\subfigure[$d_1$]
	{
		\includegraphics[width=0.16\linewidth]{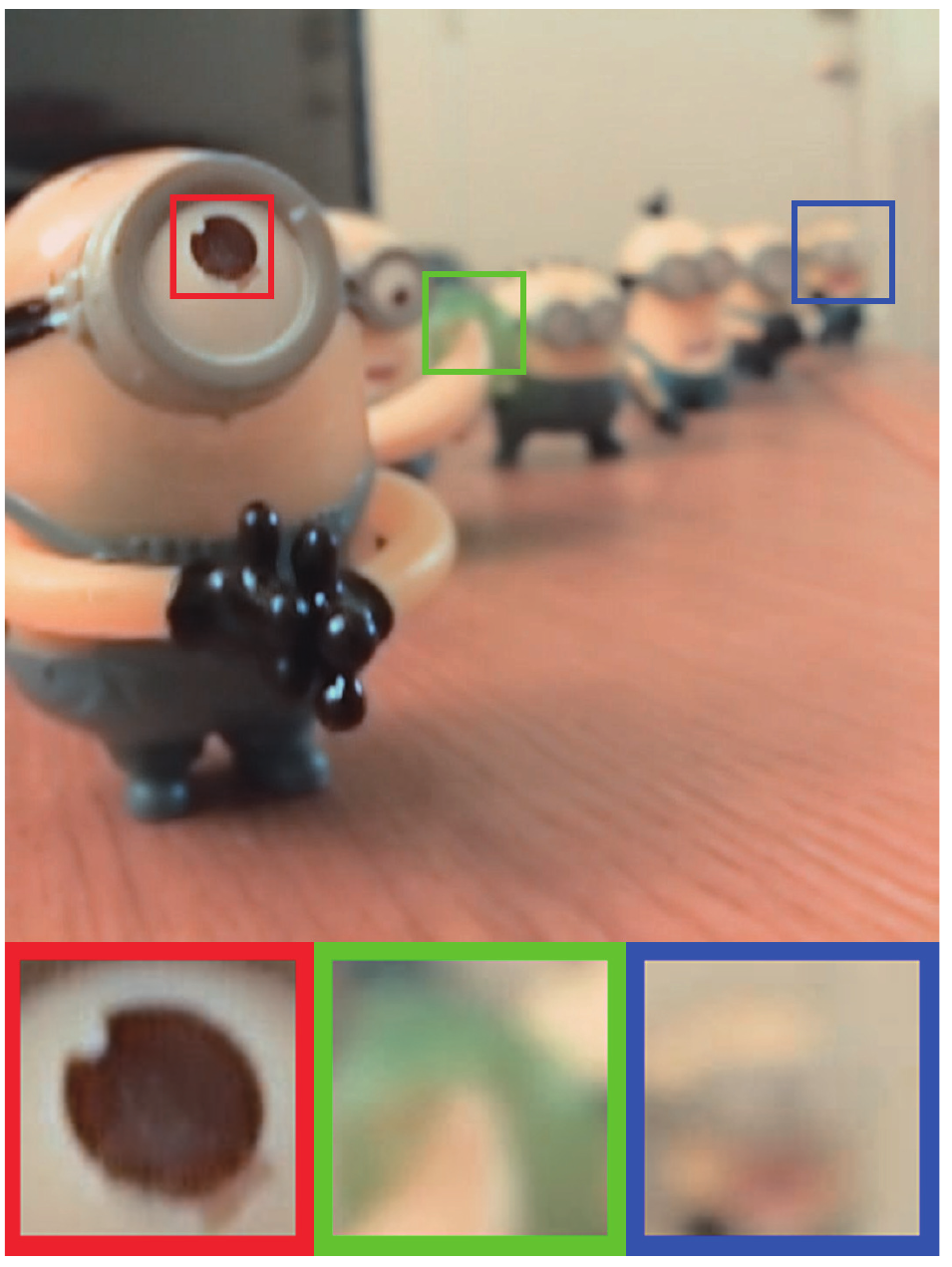}
	}
	\subfigure[$d_8$]
	{
		\includegraphics[width=0.16\linewidth]{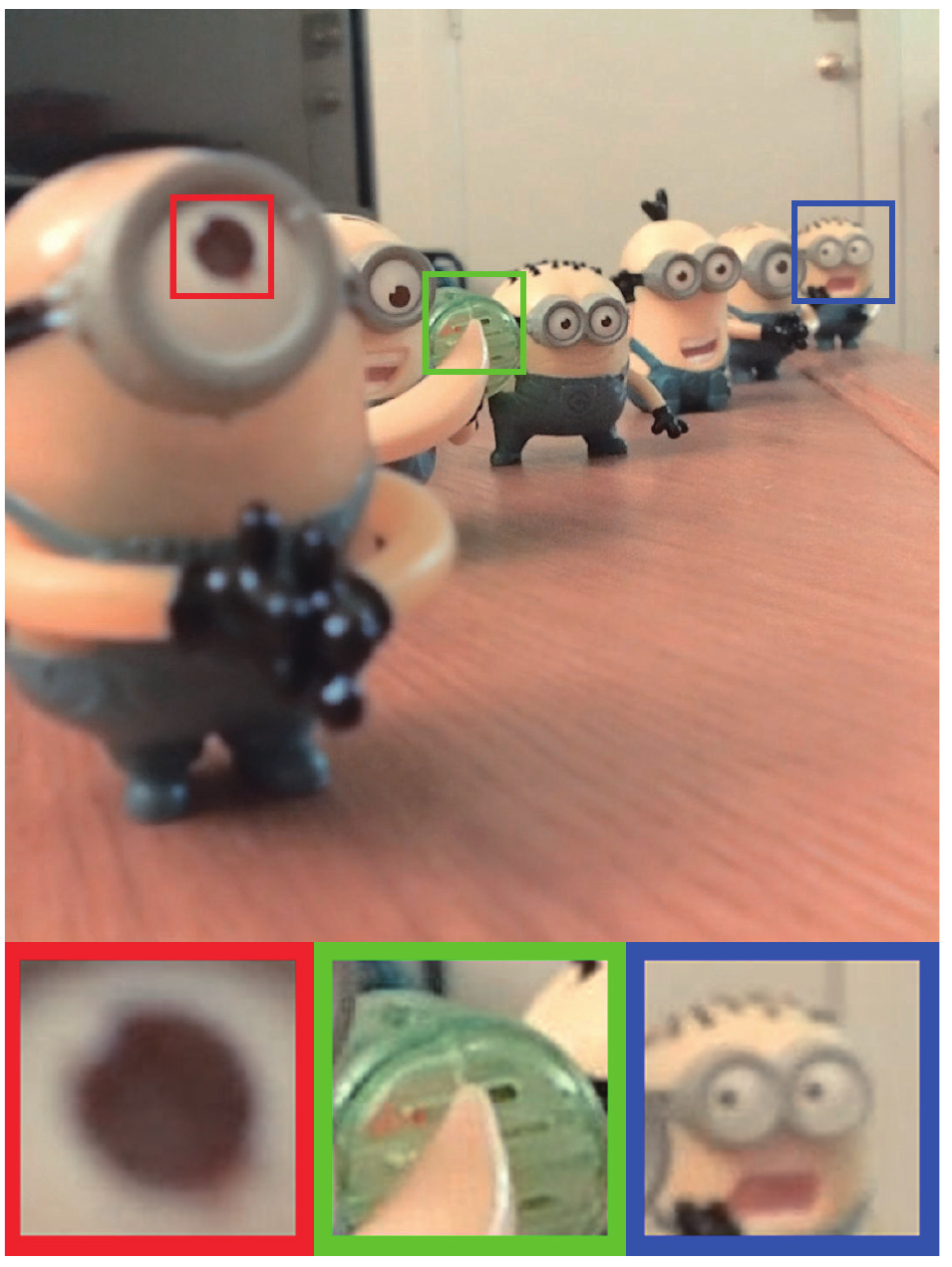}
	}
		\hspace{-2.5ex}
	\subfigure[$d_8$]
	{
		\includegraphics[width=0.16\linewidth]{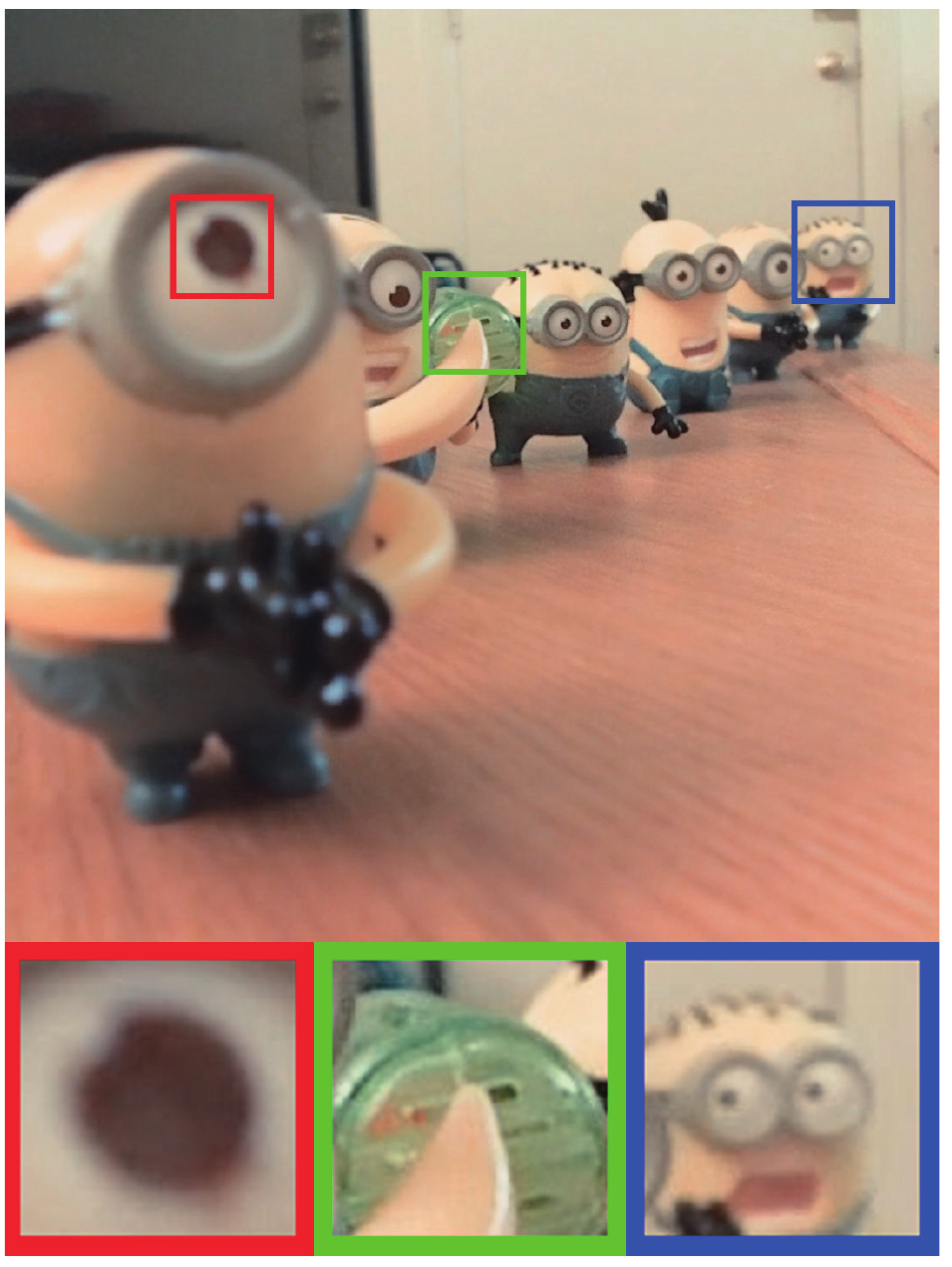}
	}
	\subfigure[$d_{10}$]
	{
		\includegraphics[width=0.16\linewidth]{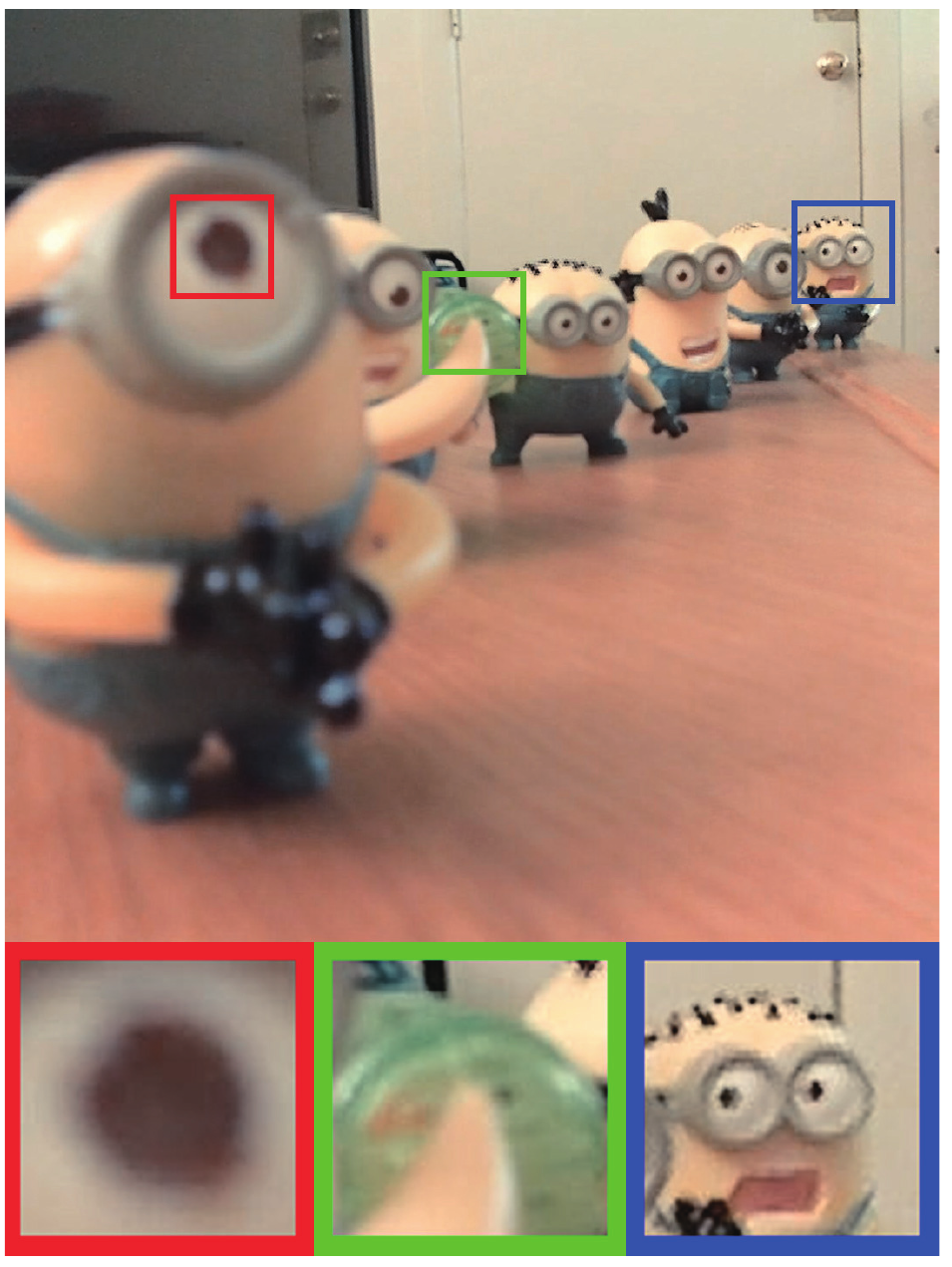}
	}
		\hspace{-2.5ex}
	\subfigure[$d_{10}$]
	{
		\includegraphics[width=0.16\linewidth]{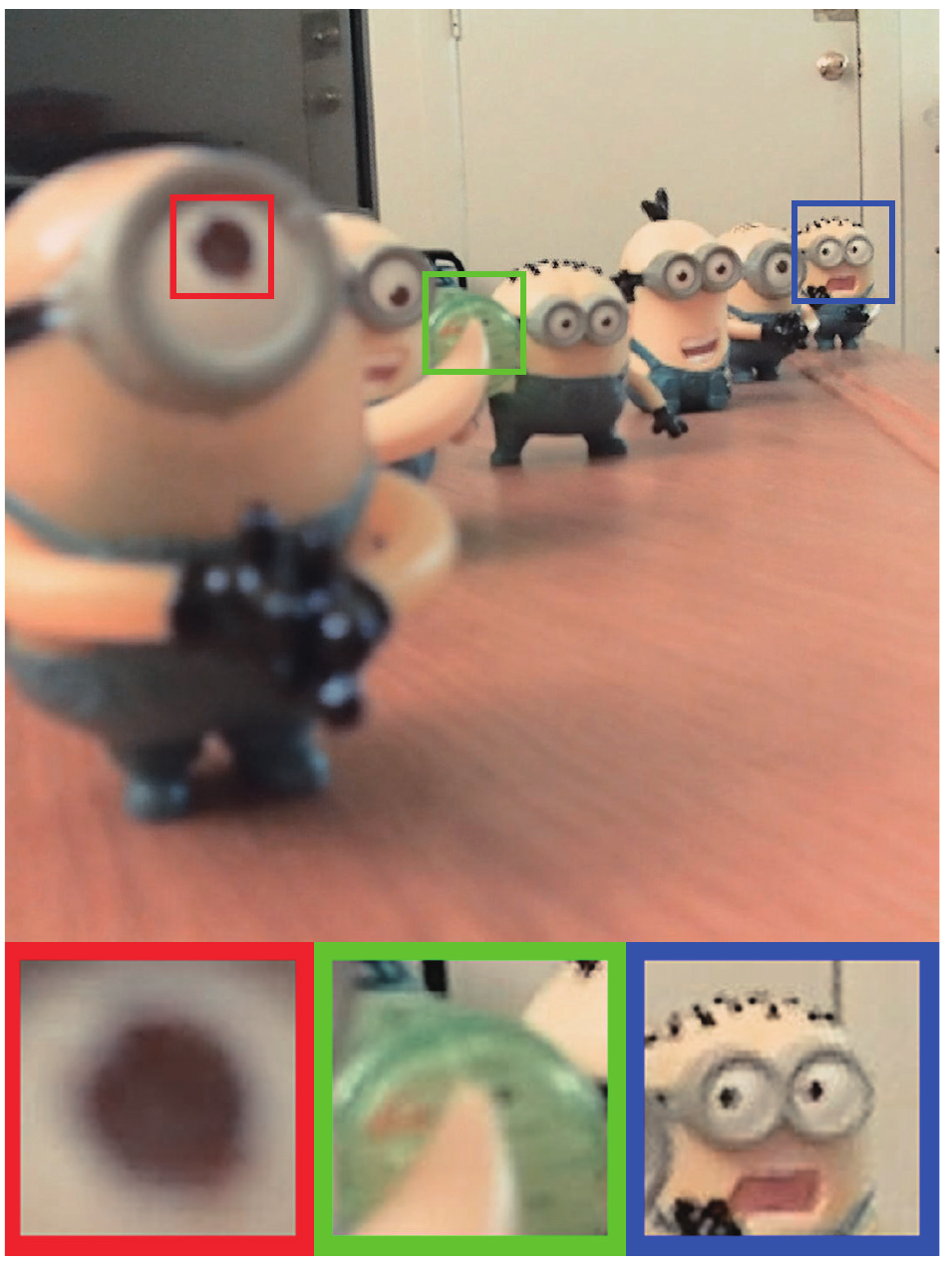}
	}
	\caption{Comparison between ground-truth images(a),(c),(e) and our restored images(b),(d),(f).}
	\label{MinionsPic}
\vspace{-3mm}
\end{figure*}

With $\textbf{A}_{i, k}$ and $\textbf{B}_{i, k}$, we can transfer the spectral information of $I_{d_i, \lambda_i}$ to $I_{d_j, \lambda_j}$. By transforming the spectral information of all the channels to a certain channel $I_{d_i, \lambda_i}$ a multispectral slice focusing on depth $d_i$ is recovered. Similarly, the rest slices of the focal stack can be reconstructed by transfering the spectral information to the rest spectral-varying slices respectively.

\section{Experimental Result}

We test the proposed chromatic aberration enlarged camera and LLT-based reconstruction algorithm on the multispectral focal stacks synthesized from both on-line dataset, i.e. LFSD \cite{li2014saliency}, and real captured images. The focal stacks in the dataset are composed of ten RGB slices focused on different depths. In order to obtain the multispectral images, we synthesize the pseudo spectra from the RGB measurement by using the training based algorithm \cite{nguyen2014training}. In our experiment, we use ten slices with spectral channels with central wavelengths 430nm, 460nm, ..., 700nm as the input, the corresponding slices are denoted by $d_1$, $d_2$, ..., $d_{10}$.  The ground truth is the full-channel focal stack composed of two multispectral slices, and each of the slices has ten spectral channels. In the experiment, we select a single channel of each multispectral slice to simulate our focal stack camera. The entire multispectral focal stacks are restored by using the proposed LLT-based reconstruction algorithm.

To quantitatively evaluate the performance, the peak signal-to-noise ratio(PSNR) and the structure similarity(SSIM) are employed. We also present reconstructed images to demonstrate the performance qualitatively. 

\begin{table}[htbp]
\vspace{-1mm}
  \caption{Quantitative evaluation for four selected depths and channels of our results.}
  \tiny
  \centering
    \begin{tabular}{ccccccccc}
    \toprule
    \multirow{2}[4]{*}{} & \multicolumn{2}{c}{$\lambda=$430nm} & \multicolumn{2}{c}{$\lambda=$520nm} & \multicolumn{2}{c}{$\lambda=$610nm} & \multicolumn{2}{c}{$\lambda=$700nm} \\
\cmidrule{2-9}          & \multicolumn{1}{c}{PSNR} & \multicolumn{1}{c}{SSIM} & \multicolumn{1}{c}{PSNR} & \multicolumn{1}{c}{SSIM} & \multicolumn{1}{c}{PSNR} & \multicolumn{1}{c}{SSIM} & \multicolumn{1}{c}{PSNR} & \multicolumn{1}{c}{SSIM} \\
    \midrule
    $d_1$     & Inf   & 1.0000  & 34.80  & 0.9799  & 35.08  & 0.9811  & 35.70  & 0.9841  \\
    \midrule
    $d_4$     & 39.79  & 0.9768  & Inf   & 1.0000  & 37.76  & 0.9829  & 35.26  & 0.9729  \\
    \midrule
    $d_7$     & 42.27  & 0.9867  & 36.28  & 0.9808  & Inf   & 1.0000  & 42.26  & 0.9959  \\
    \midrule
    $d_{10}$    & 38.20  & 0.9801  & 30.44  & 0.9684  & 33.55  & 0.9859  & Inf   & 1.0000  \\
    \bottomrule
    \end{tabular}%
  \label{MinionsTab}%
\vspace{-2mm}
\end{table}

\paragraph{Quantitative evaluation.} The average quantitative measurements of restored images are shown in Tab.~\ref{MinionsTab}. It is obvious that the proposed method can achieve promising performance in terms of both PSNR and SSIM metrics.

\begin{figure}[h]
	\centering
	\begin{minipage}[c]{1.0\linewidth}
	\scriptsize\rotatebox{90}{$\lambda = 430$nm}
		\includegraphics[width=0.23\linewidth]{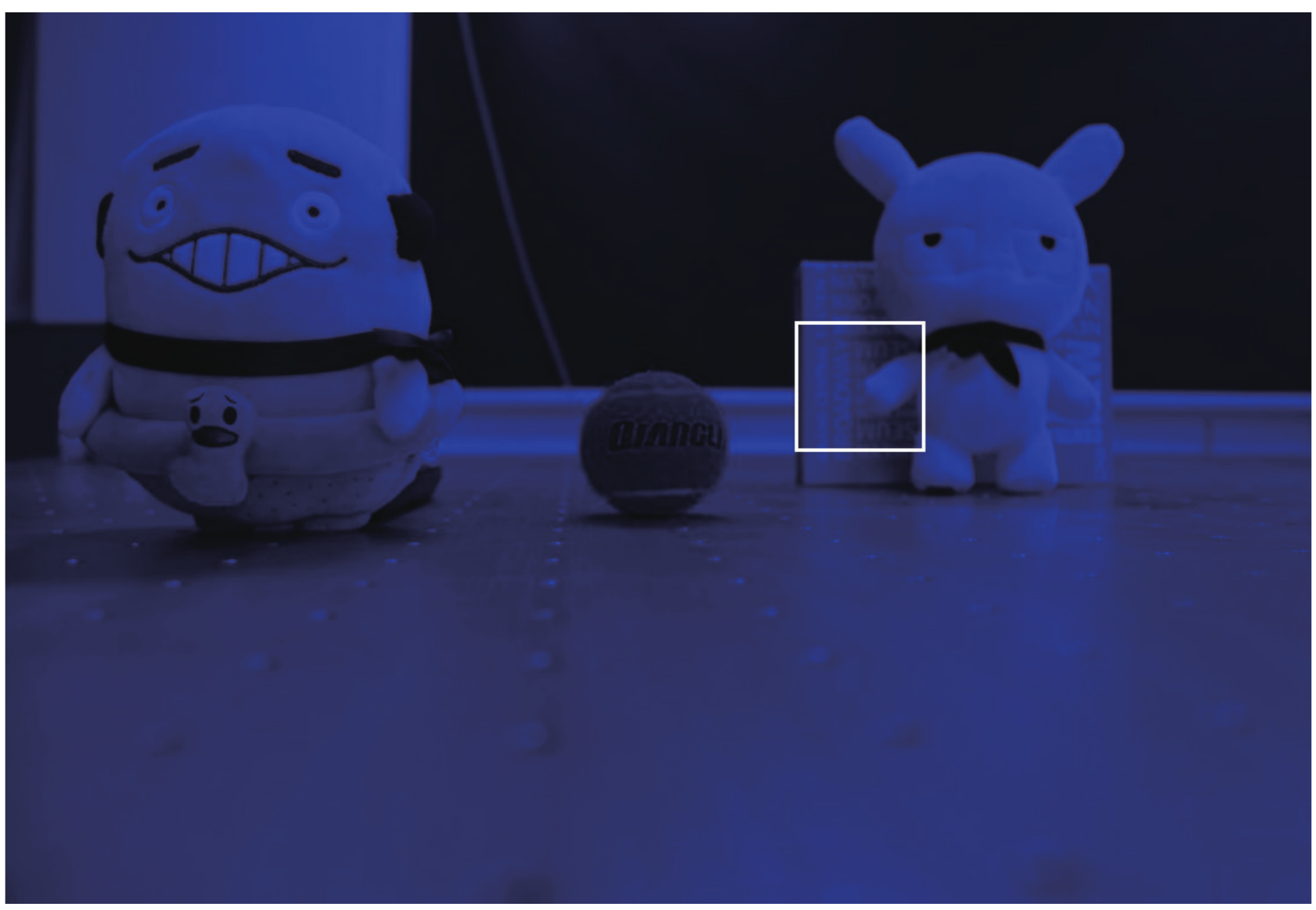}
\hfill
		\includegraphics[width=0.23\linewidth]{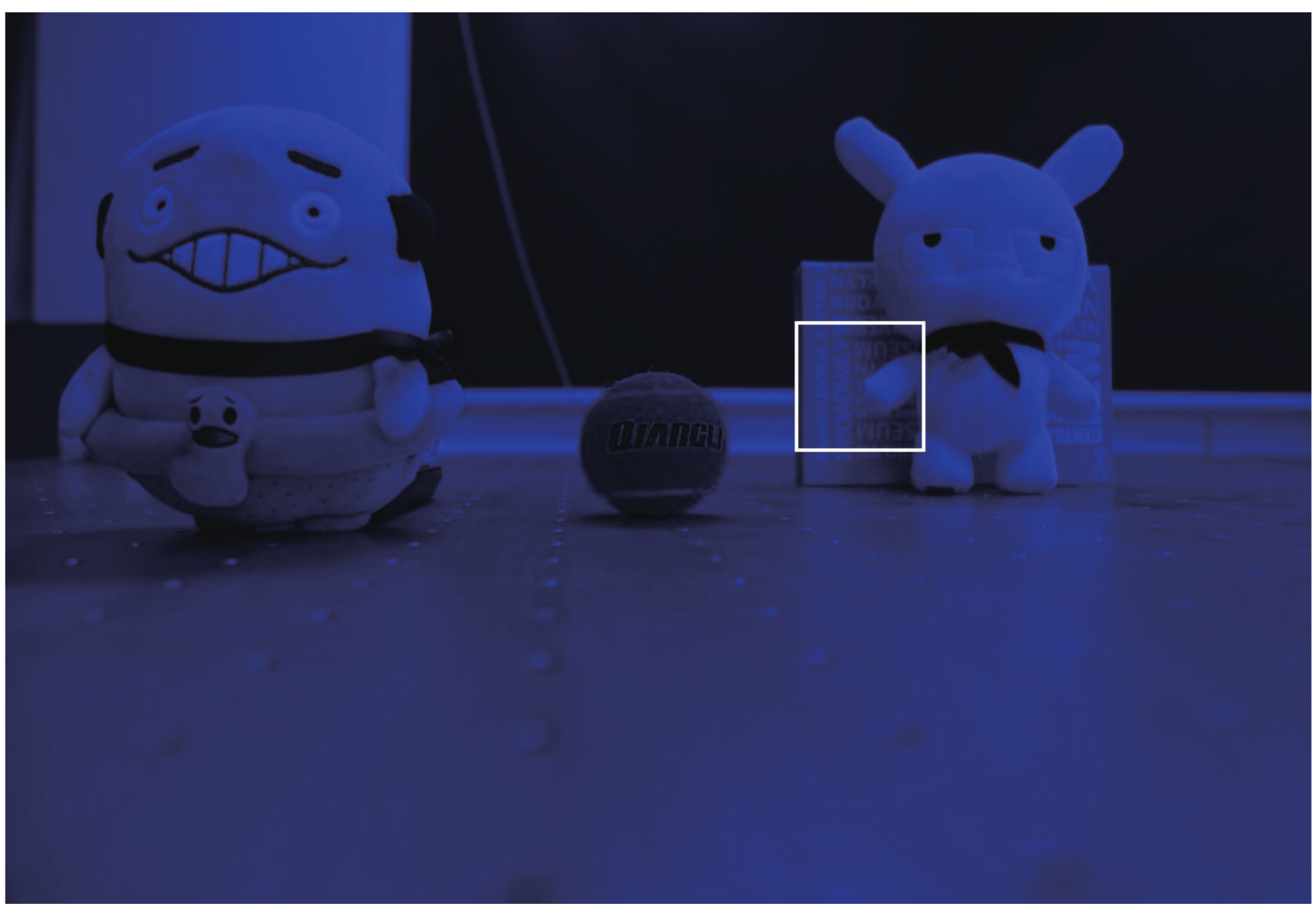}
\hfill
		\includegraphics[width=0.23\linewidth]{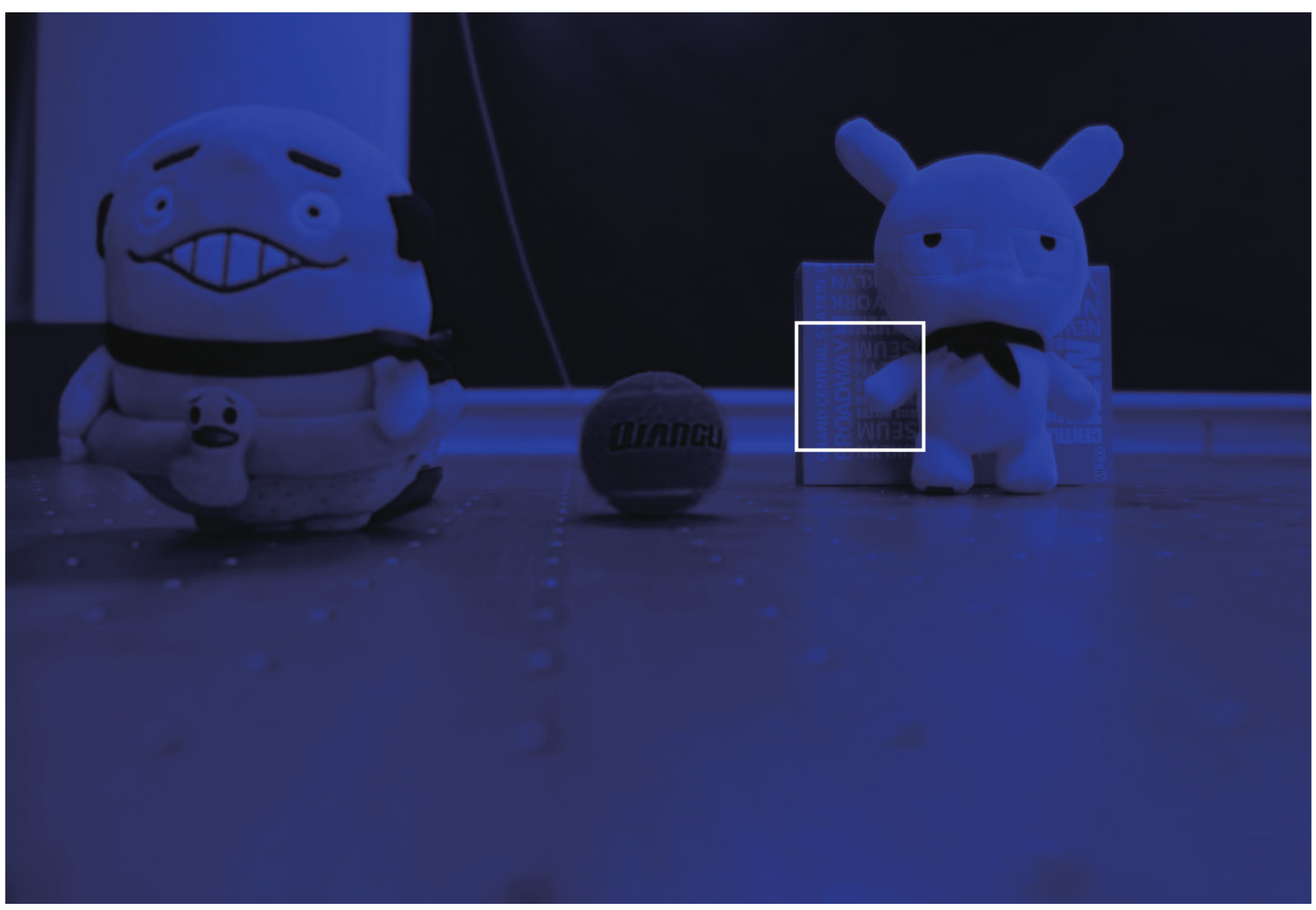}
\hfill
		\includegraphics[width=0.23\linewidth]{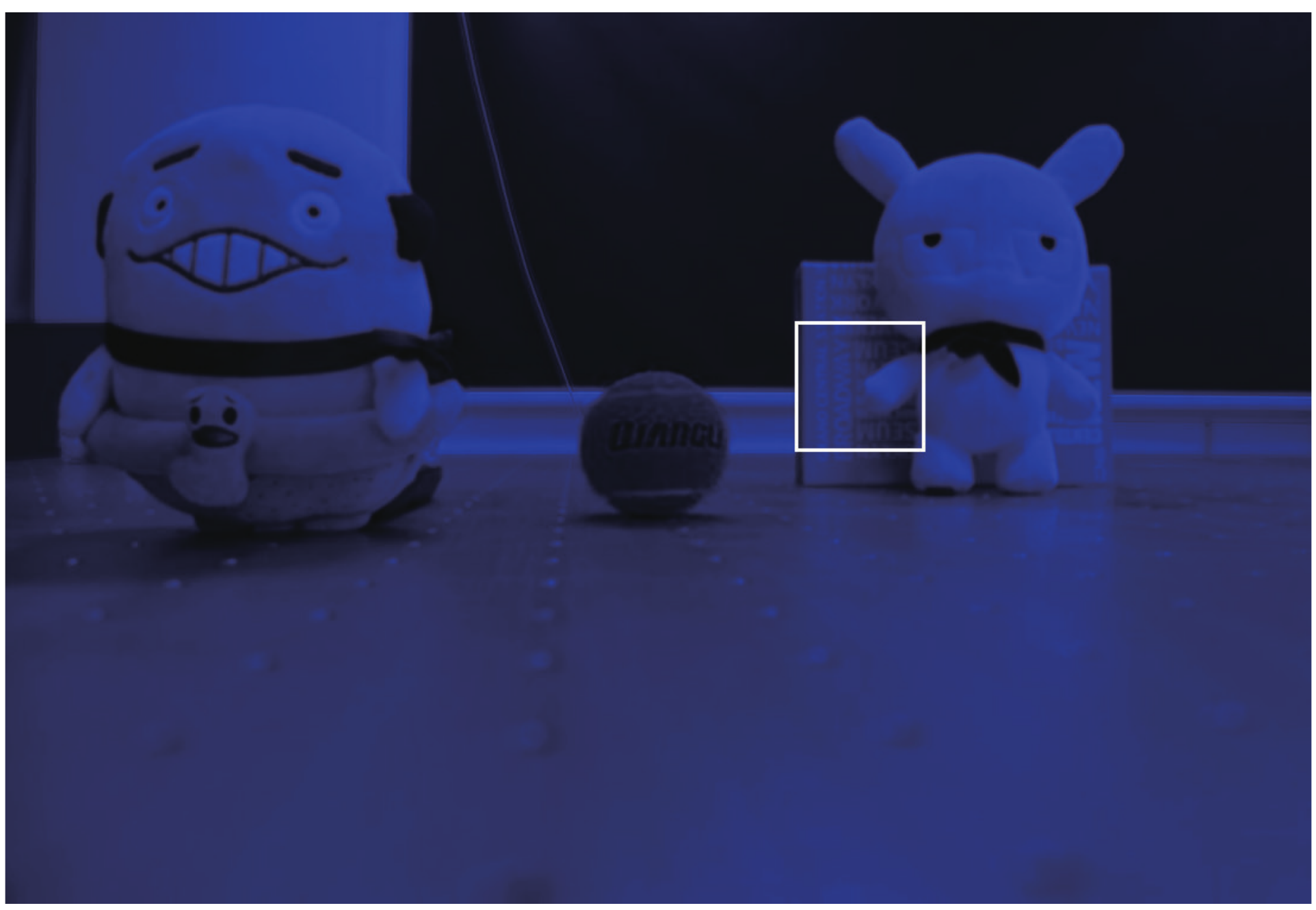}
	
	\end{minipage}
	\begin{minipage}[c]{1.0\linewidth}
	\scriptsize\rotatebox{90}{$\lambda = 520$nm}
		\includegraphics[width=0.23\linewidth]{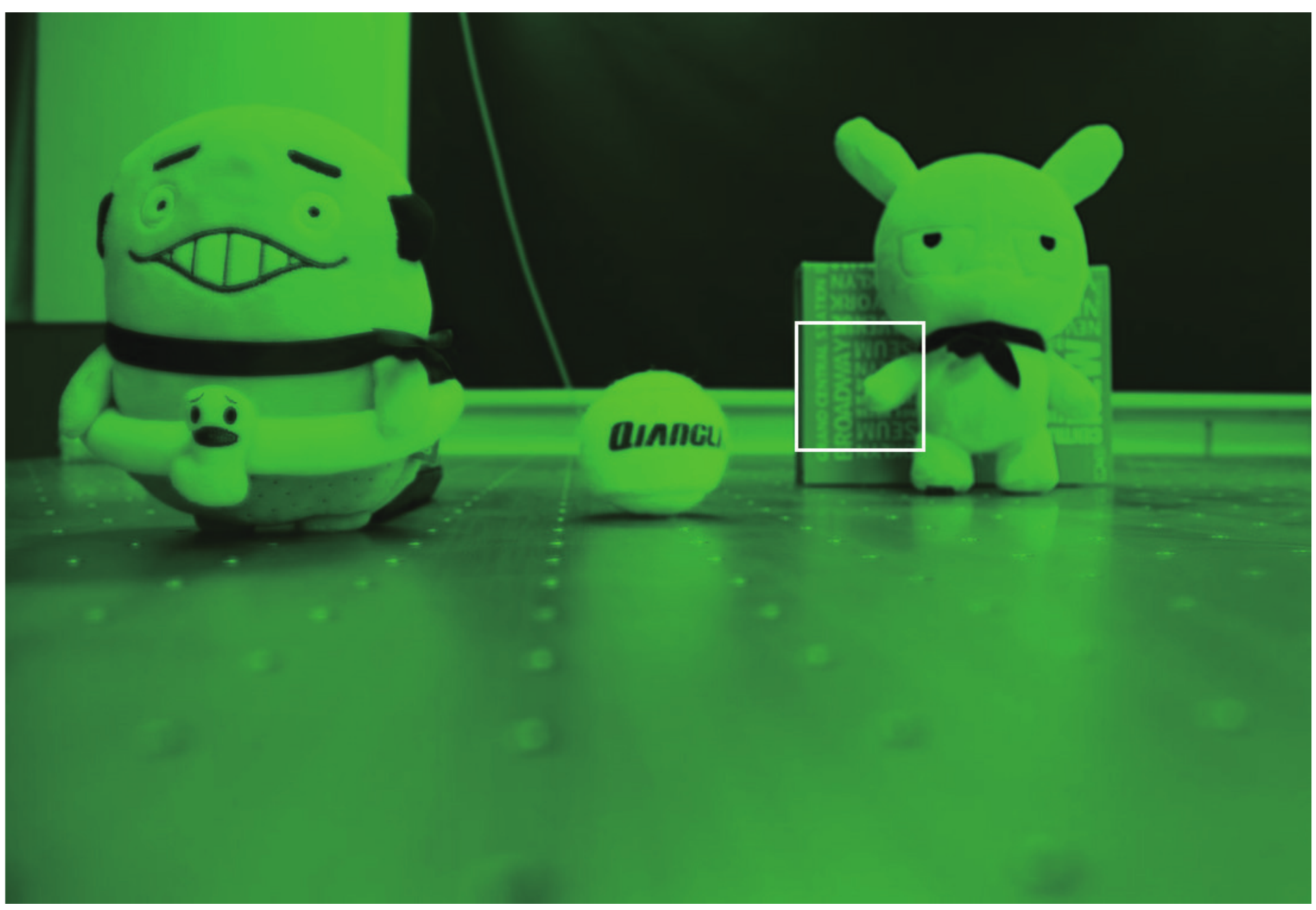}
\hfill
		\includegraphics[width=0.23\linewidth]{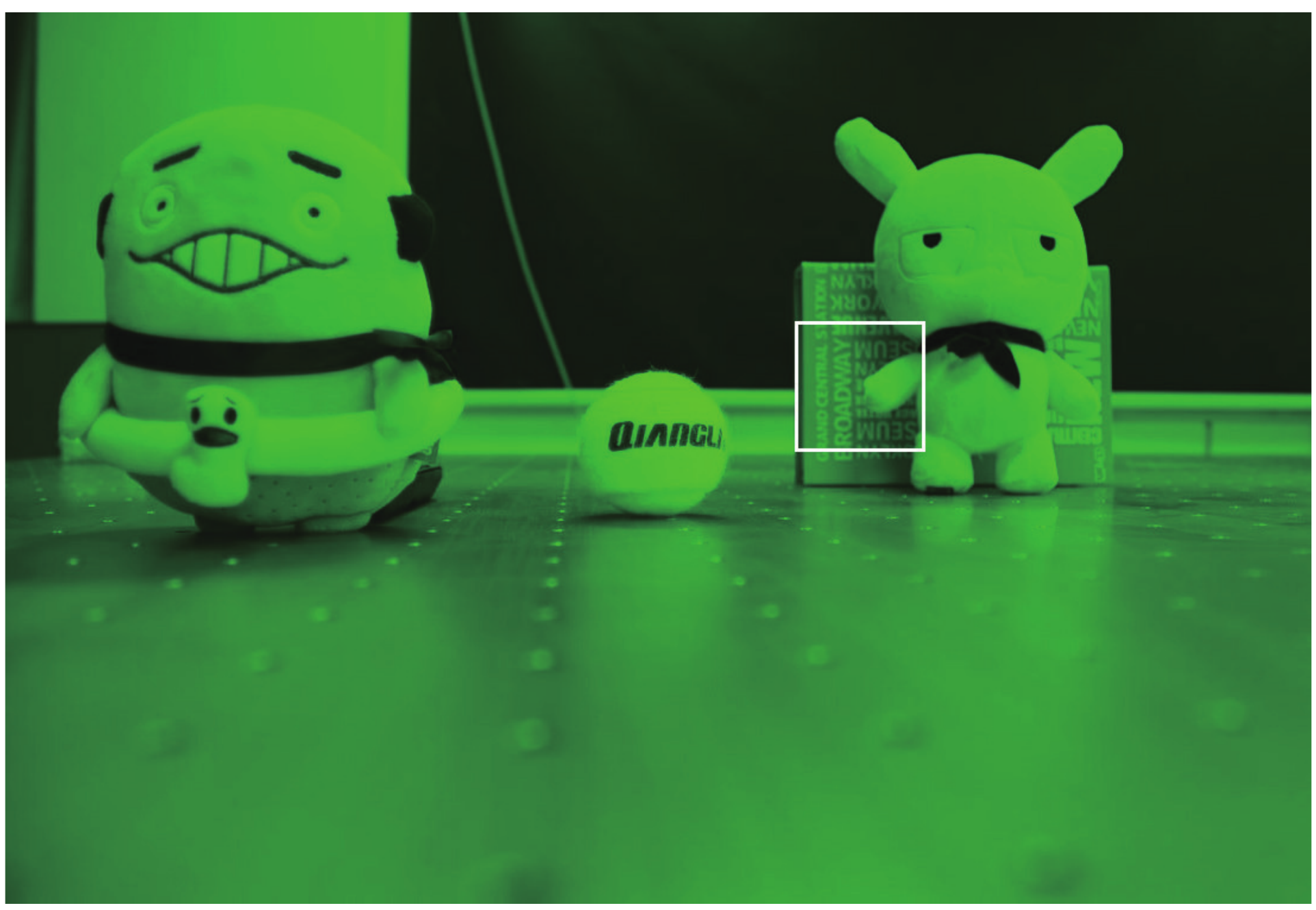}
\hfill	
		\includegraphics[width=0.23\linewidth]{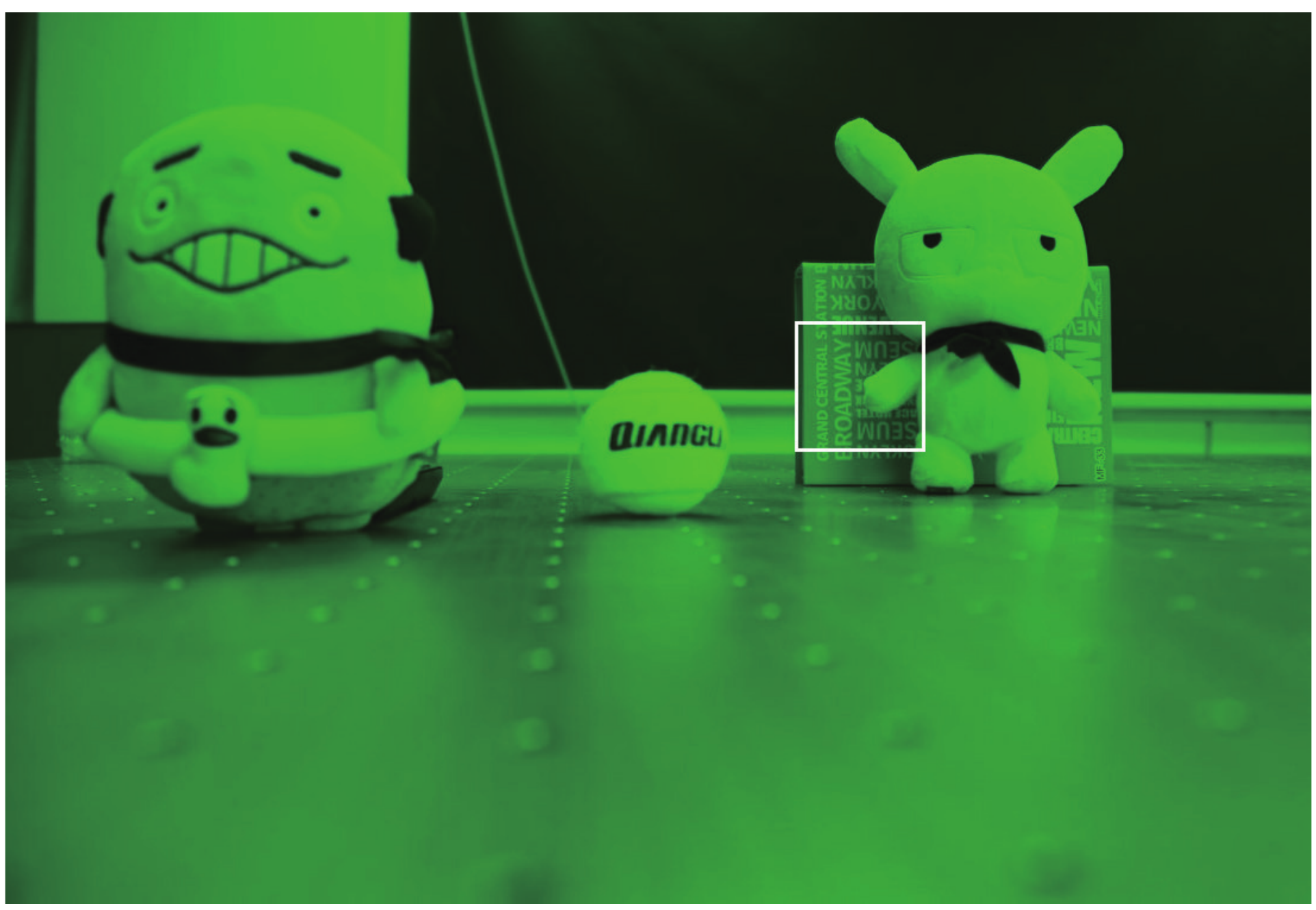}
\hfill	
		\includegraphics[width=0.23\linewidth]{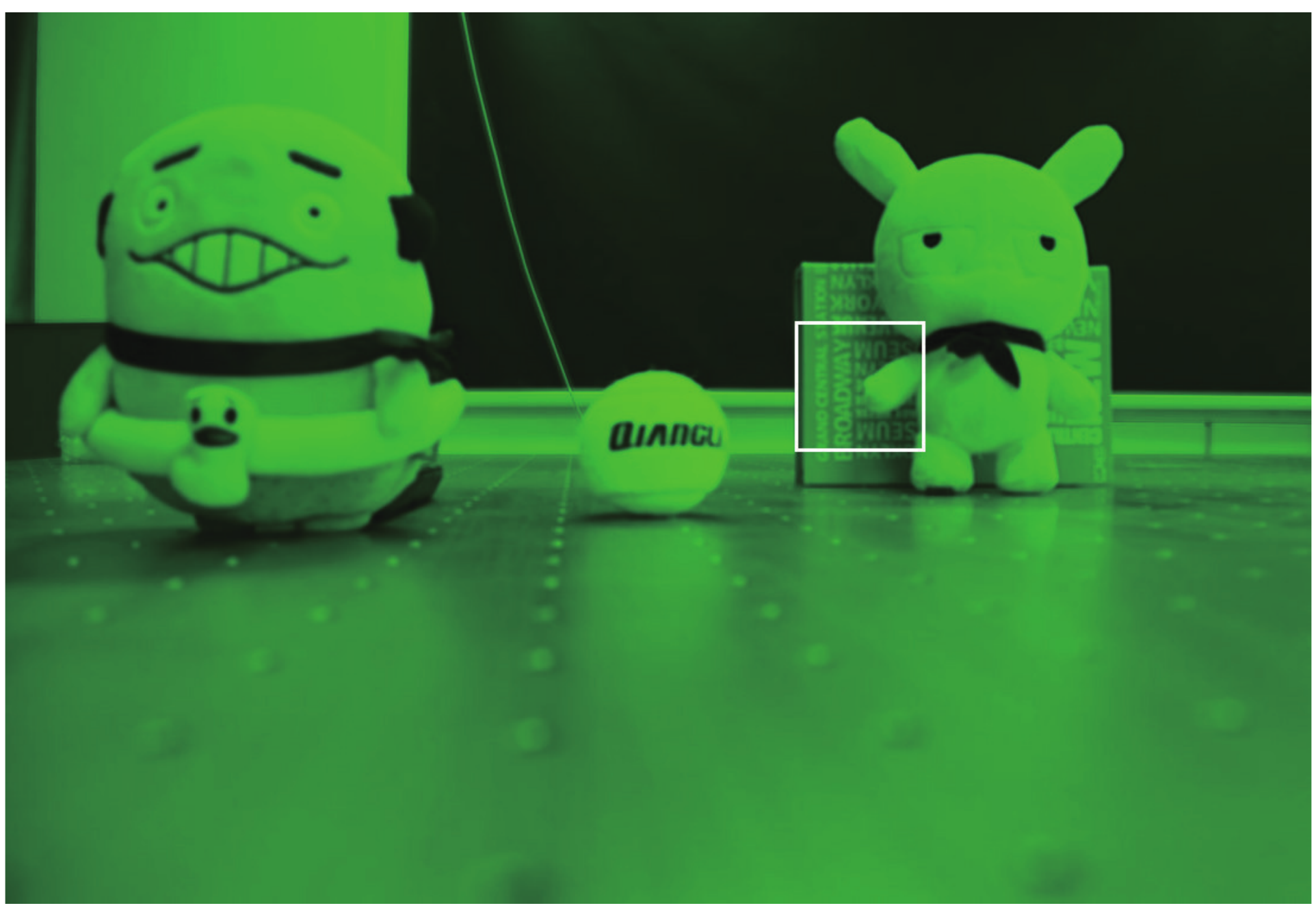}
	
	\end{minipage}
	\begin{minipage}[c]{1.0\linewidth}
	\scriptsize\rotatebox{90}{$\lambda = 610$nm}
		\includegraphics[width=0.23\linewidth]{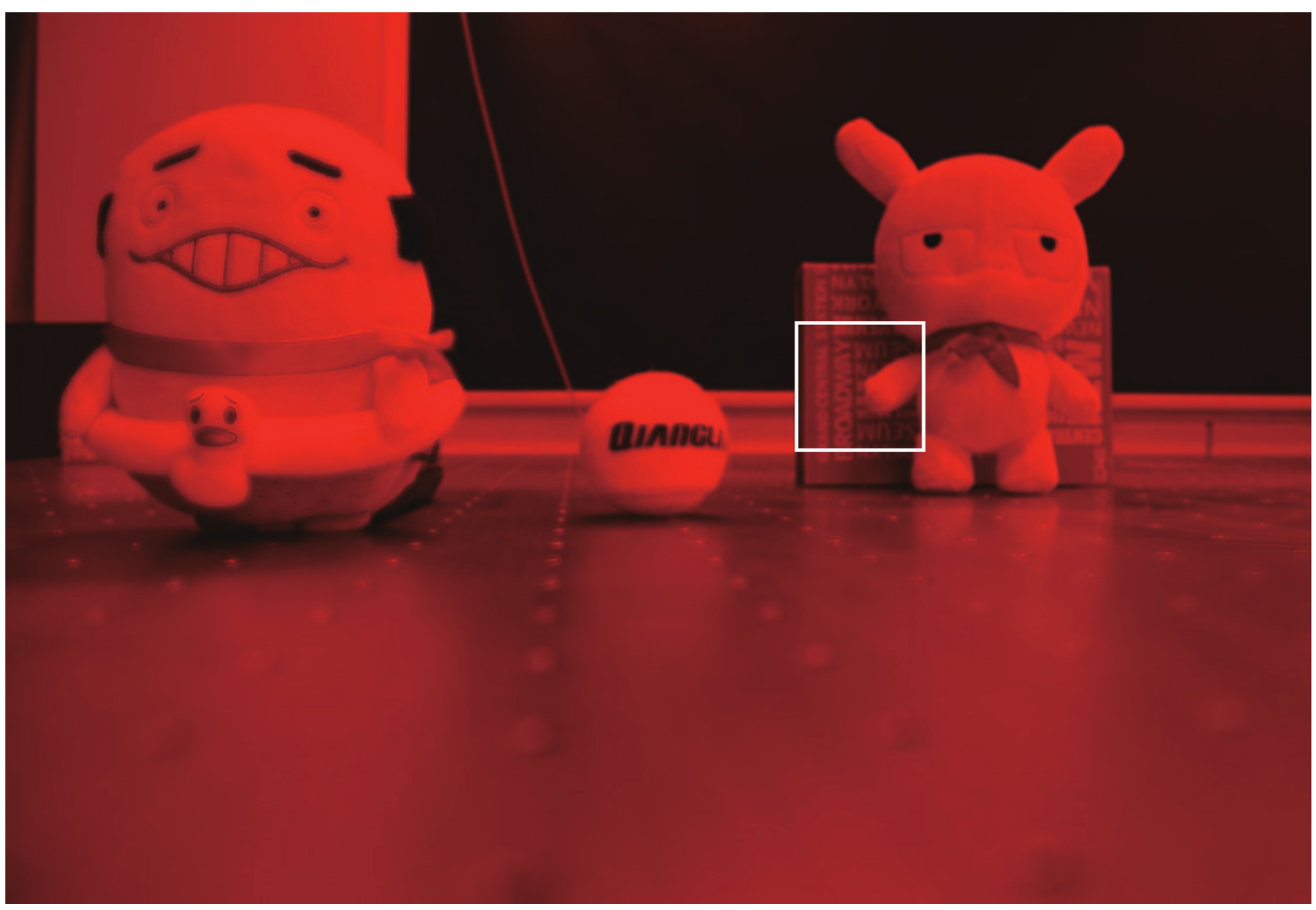}
\hfill	
		\includegraphics[width=0.23\linewidth]{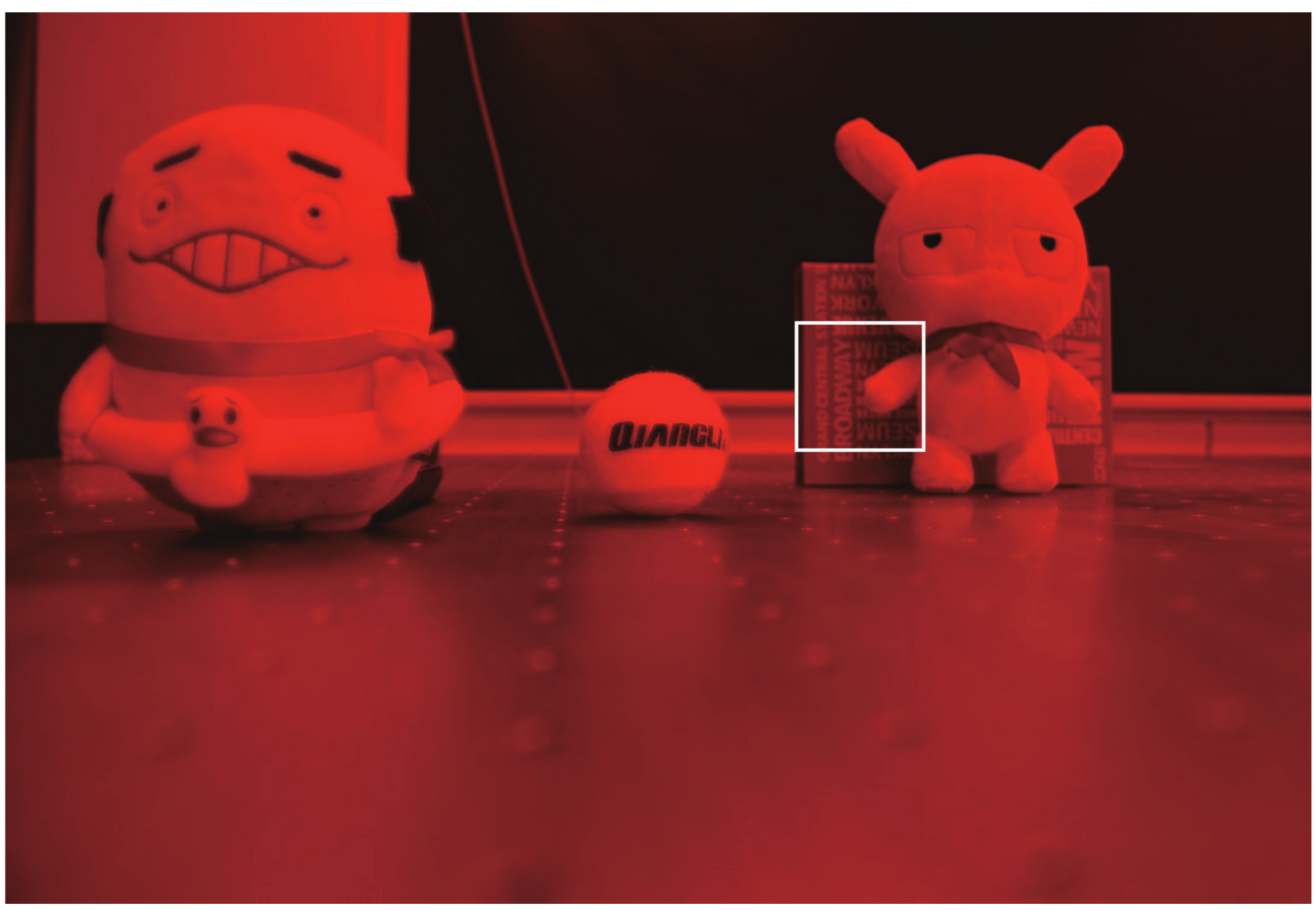}
\hfill	
		\includegraphics[width=0.23\linewidth]{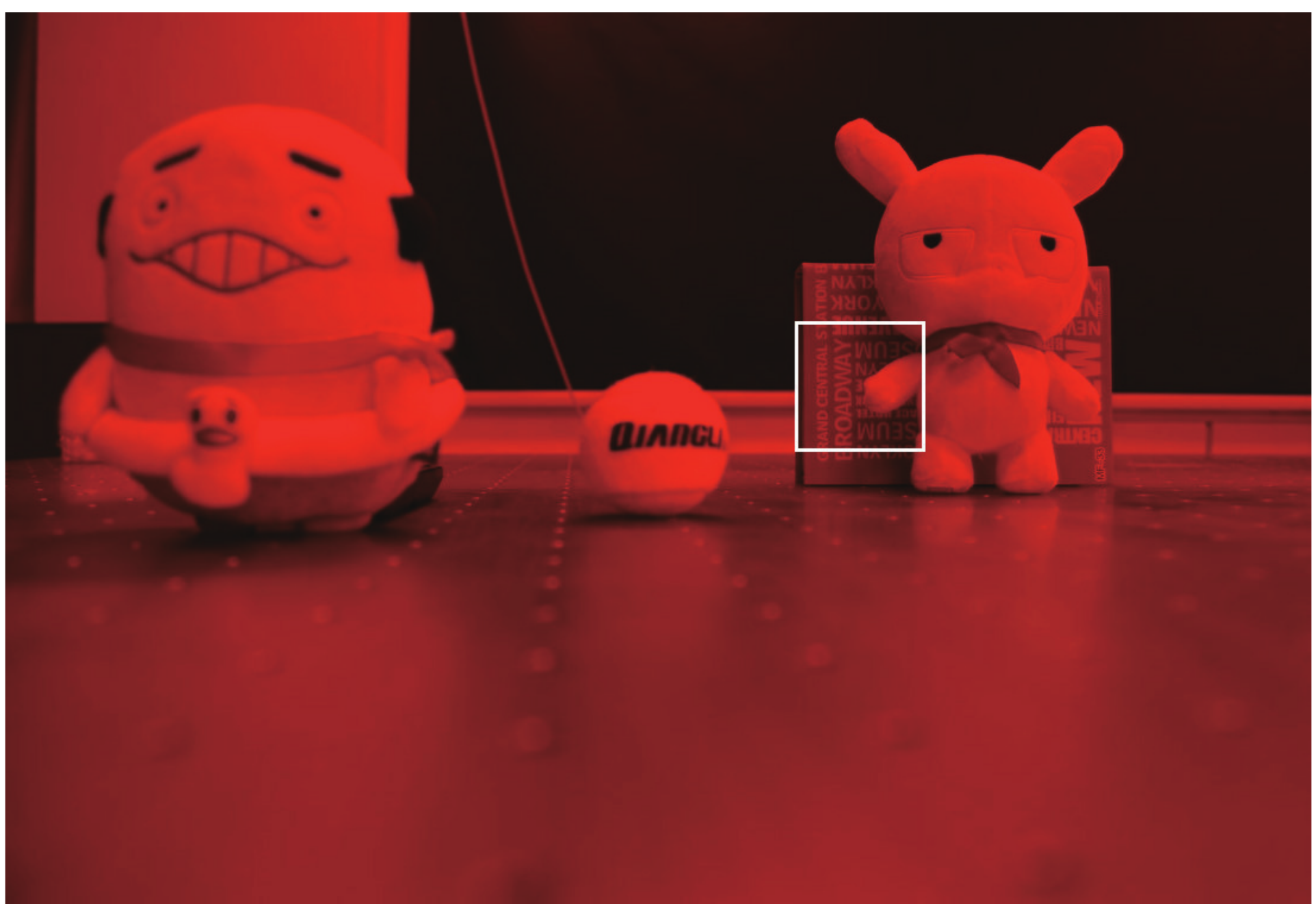}
\hfill	
		\includegraphics[width=0.23\linewidth]{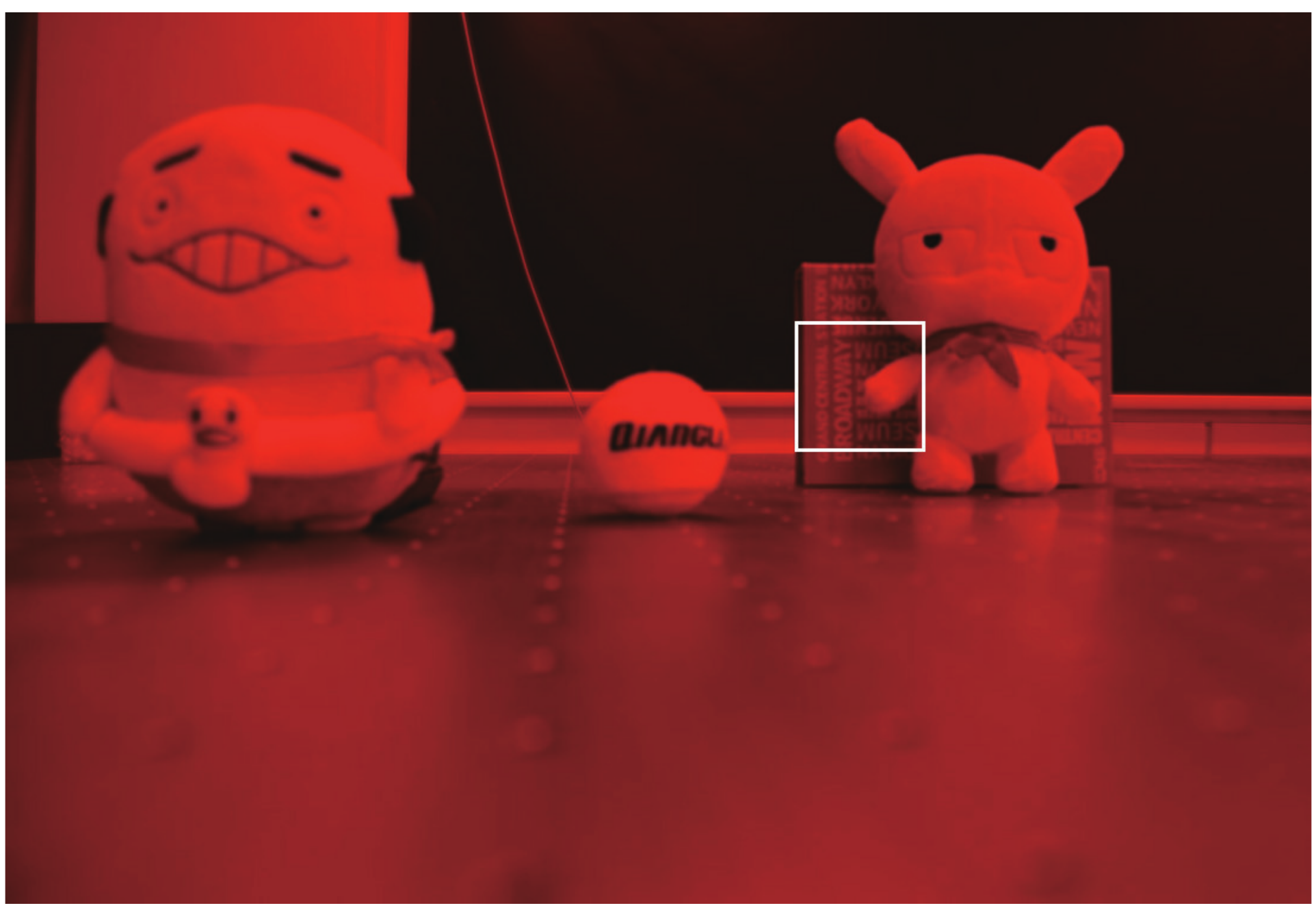}
	
	\end{minipage}
\vspace{0.25cm}
	\begin{minipage}[c]{1.0\linewidth}
	\scriptsize\rotatebox{90}{$\lambda = 700$nm}
		\includegraphics[width=0.23\linewidth]{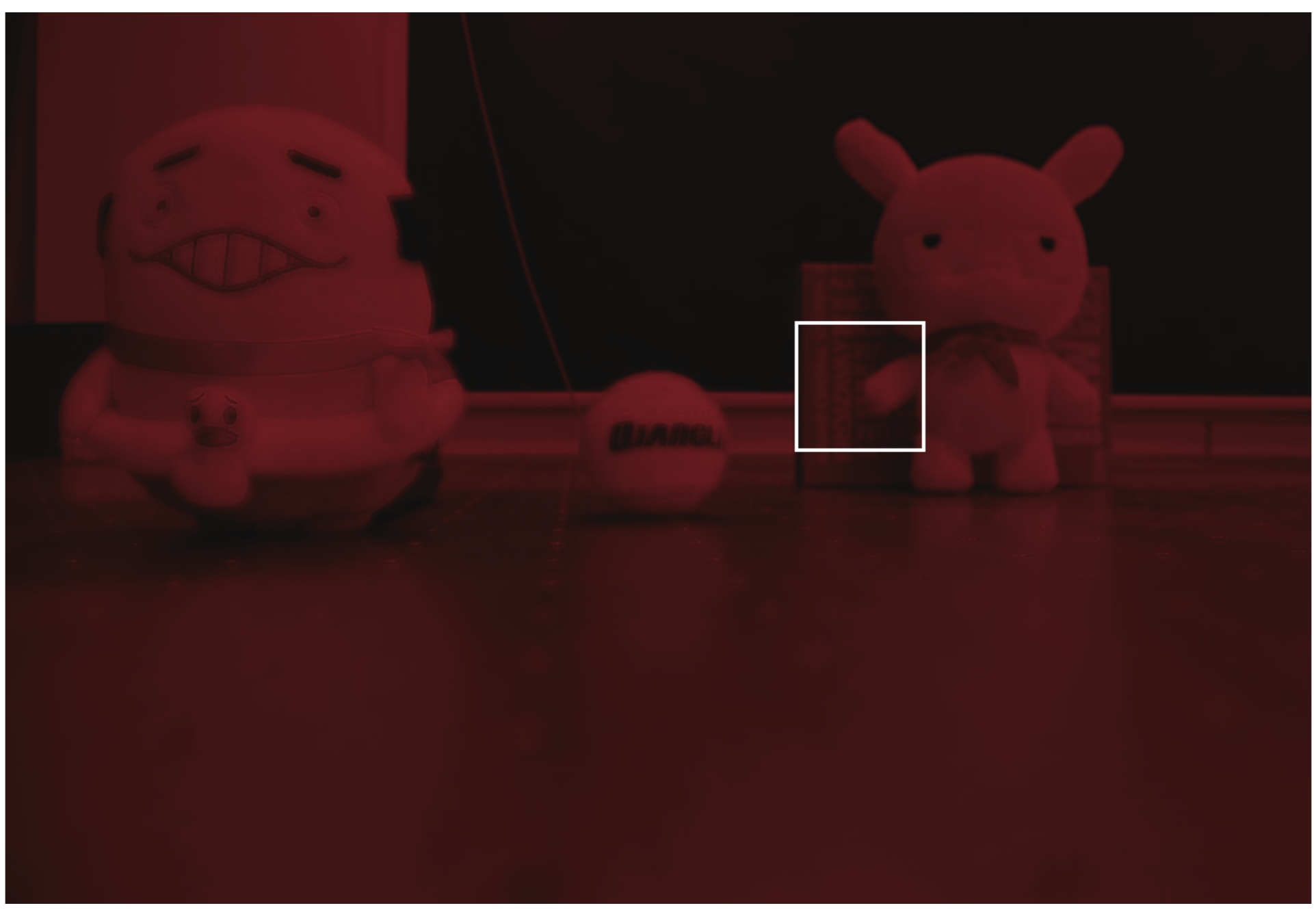}
\hfill	
		\includegraphics[width=0.23\linewidth]{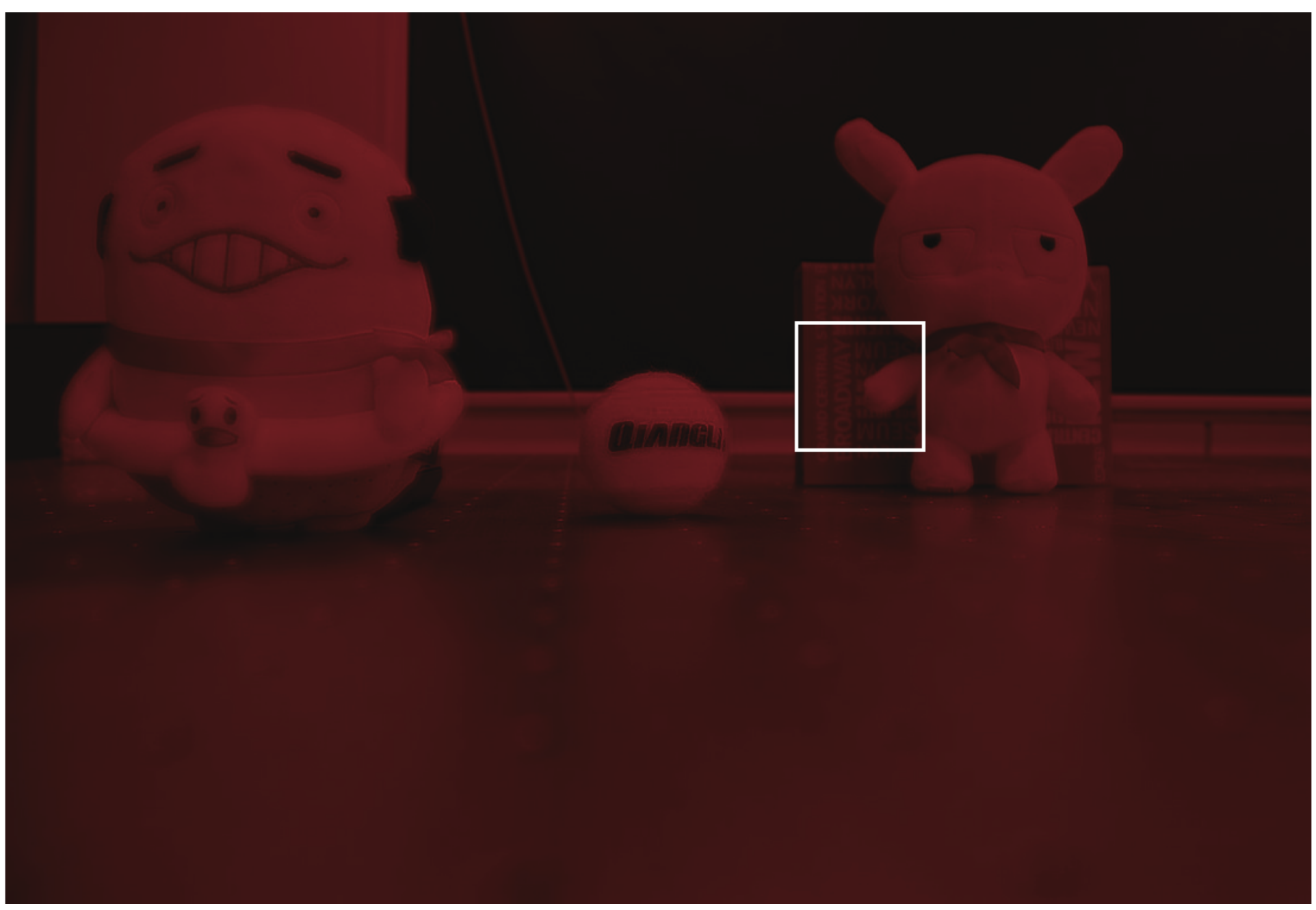}
\hfill	
		\includegraphics[width=0.23\linewidth]{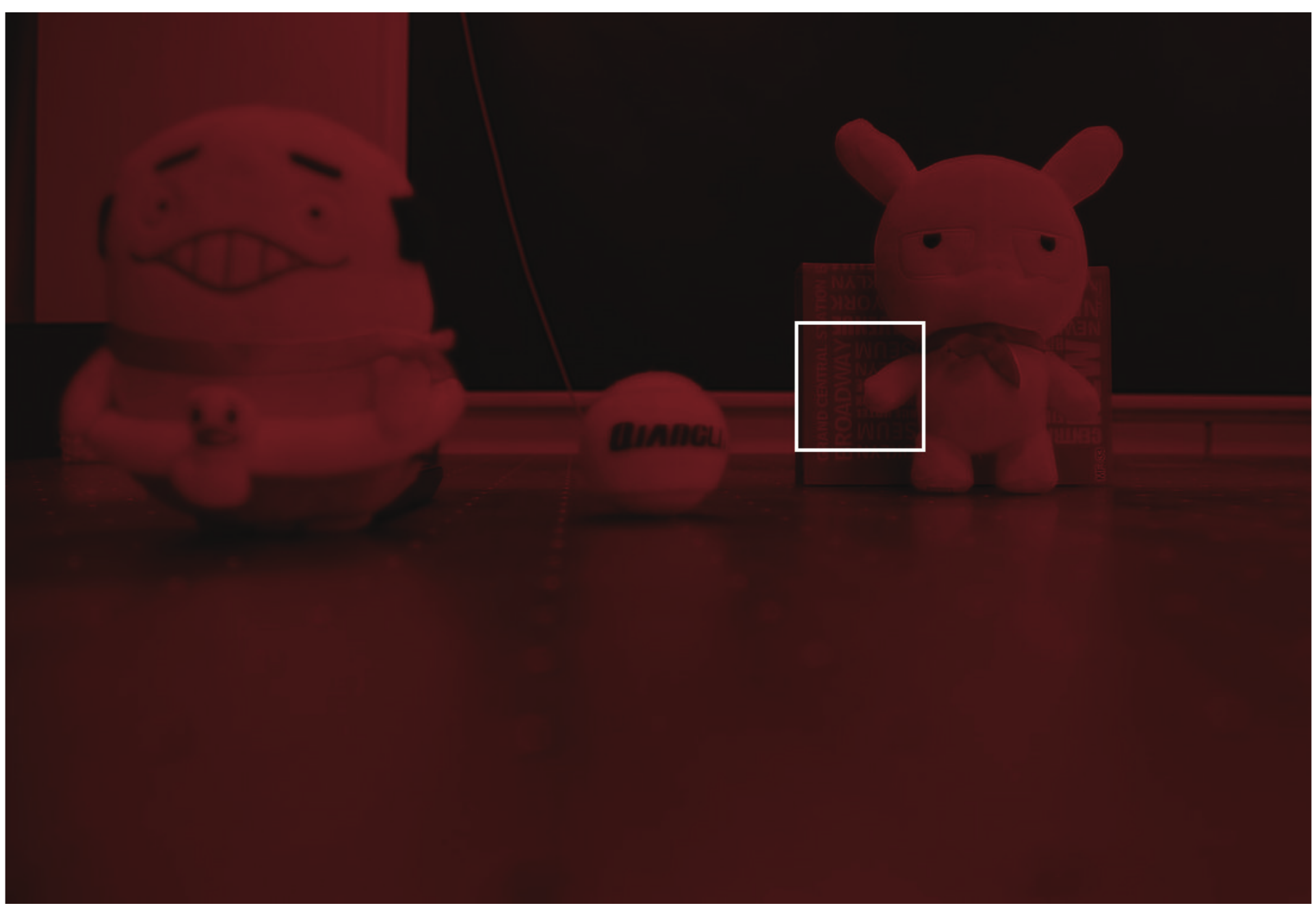}
\hfill	
		\includegraphics[width=0.23\linewidth]{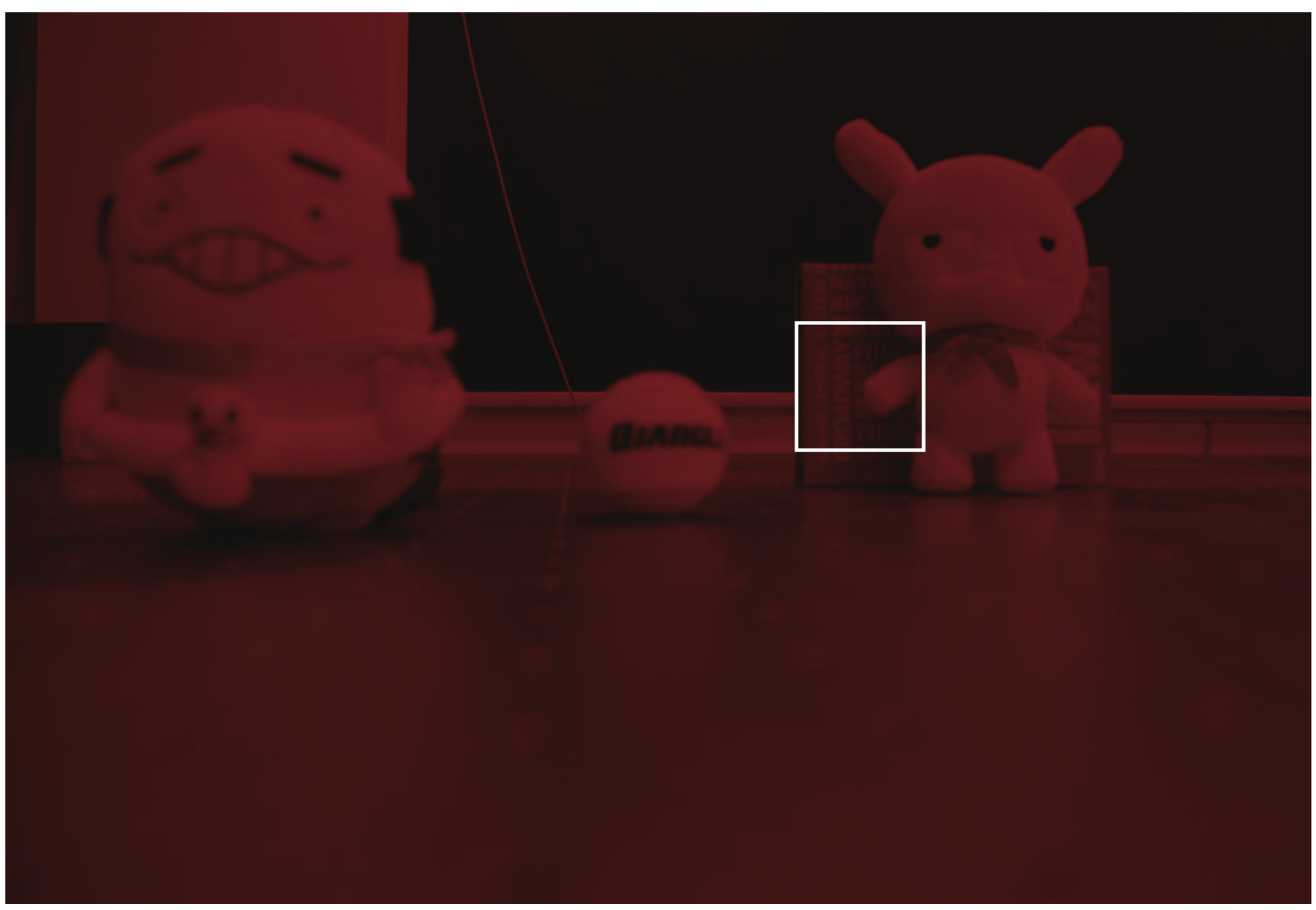}
	
	\end{minipage}
	\begin{minipage}[c]{1.0\linewidth}
	\scriptsize\rotatebox{90}{Details of results}
		\includegraphics[width=0.23\linewidth]{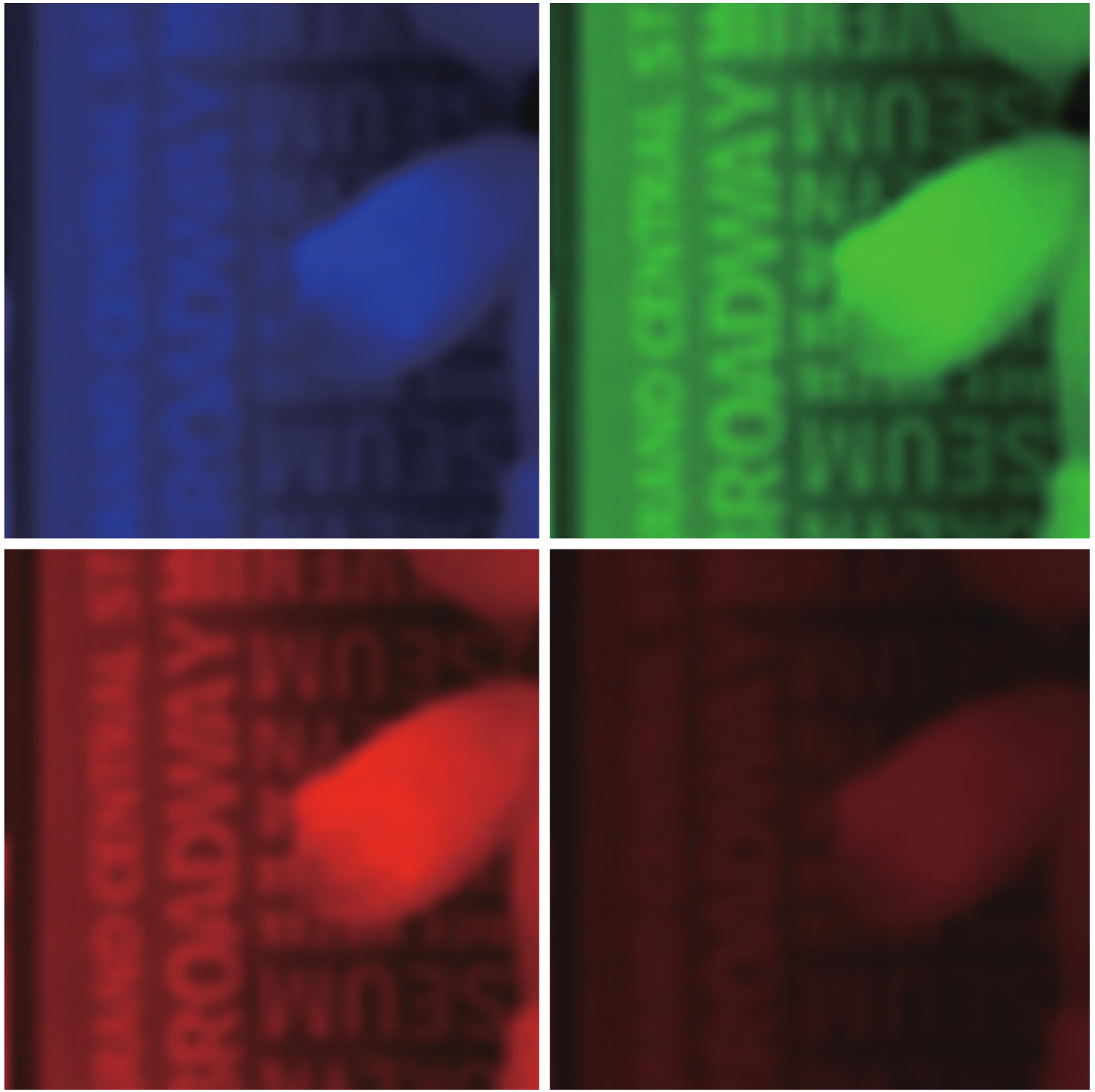}
\hfill	
		\includegraphics[width=0.23\linewidth]{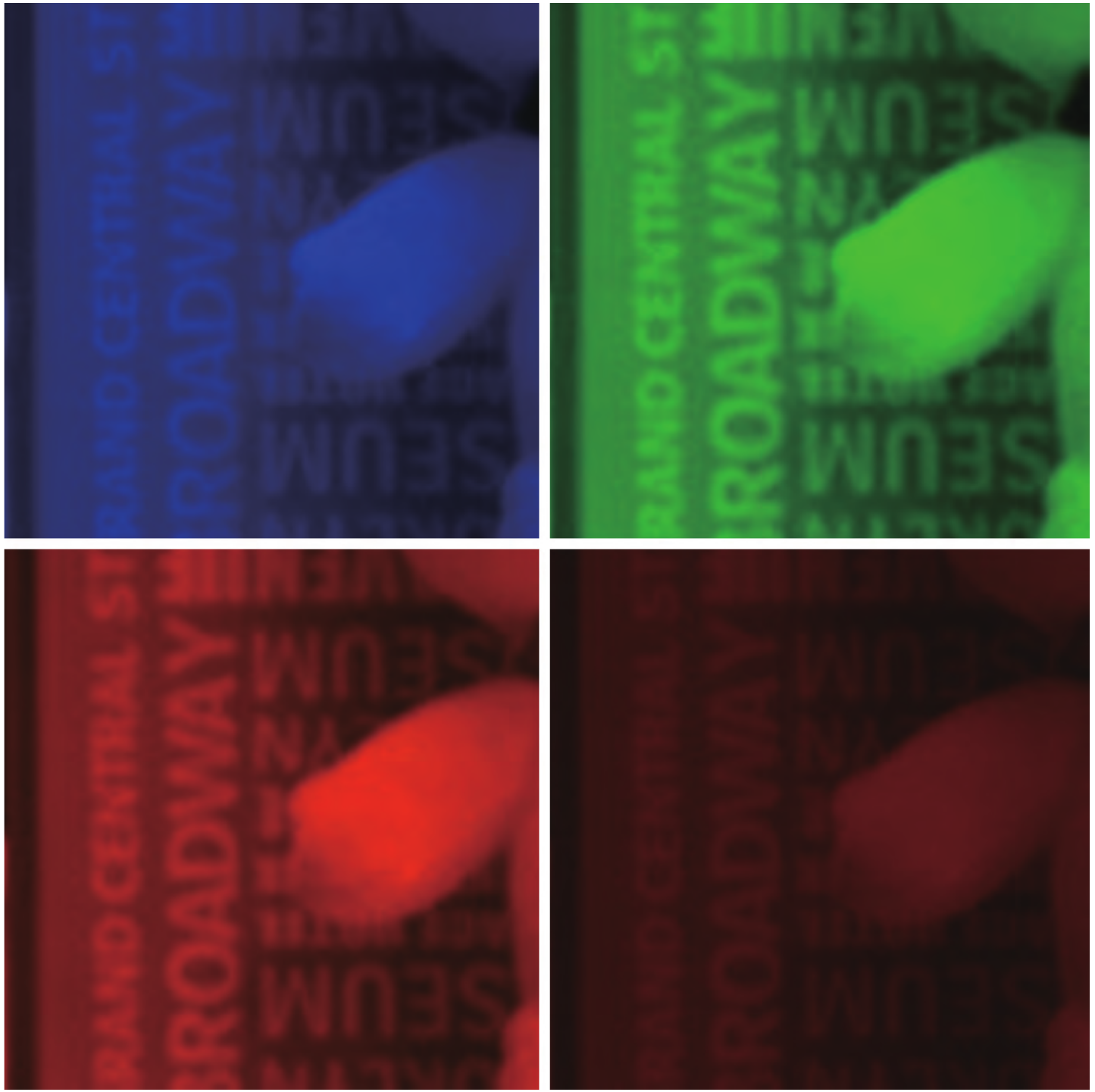}
\hfill	
		\includegraphics[width=0.23\linewidth]{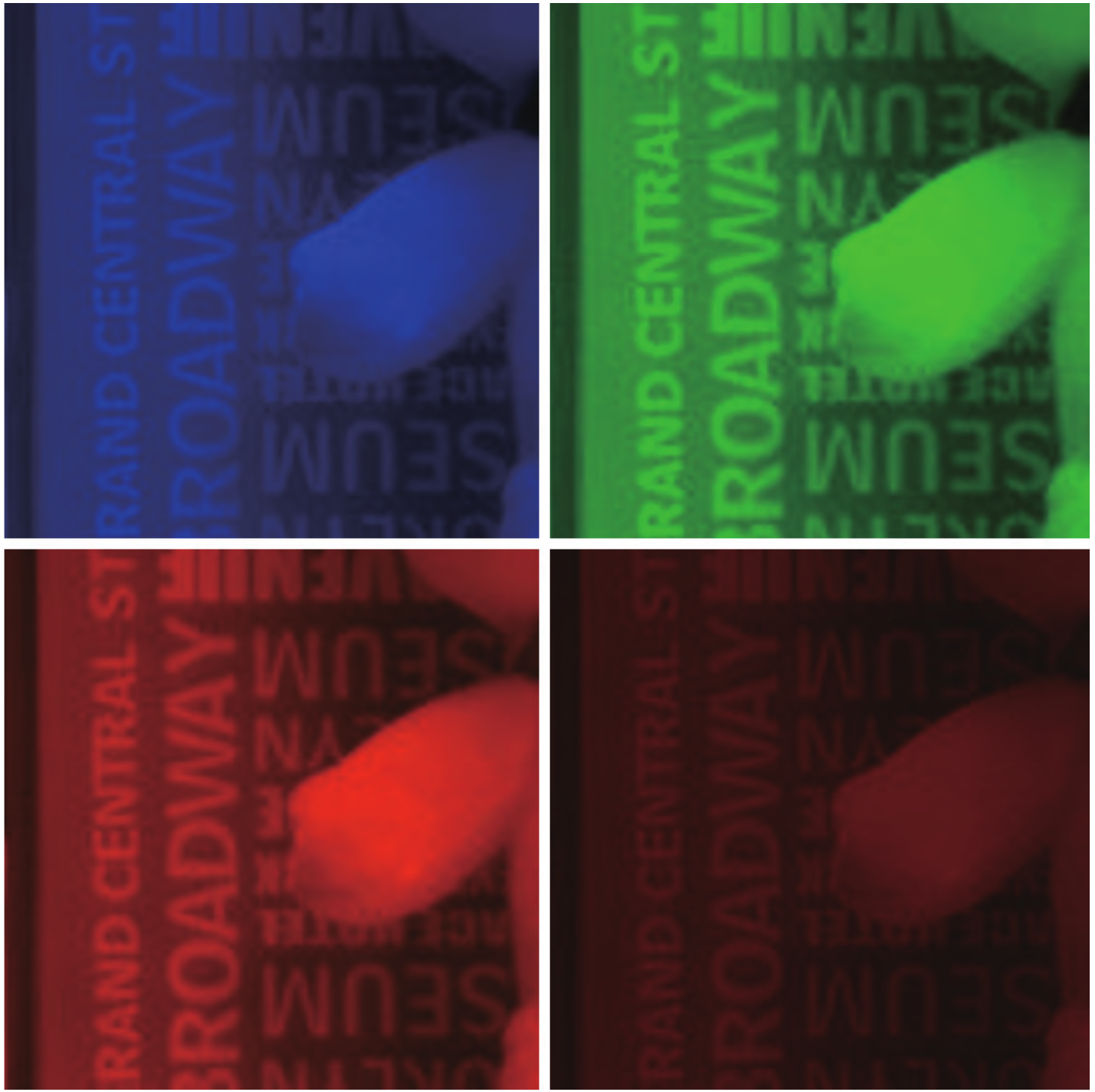}
\hfill	
		\includegraphics[width=0.23\linewidth]{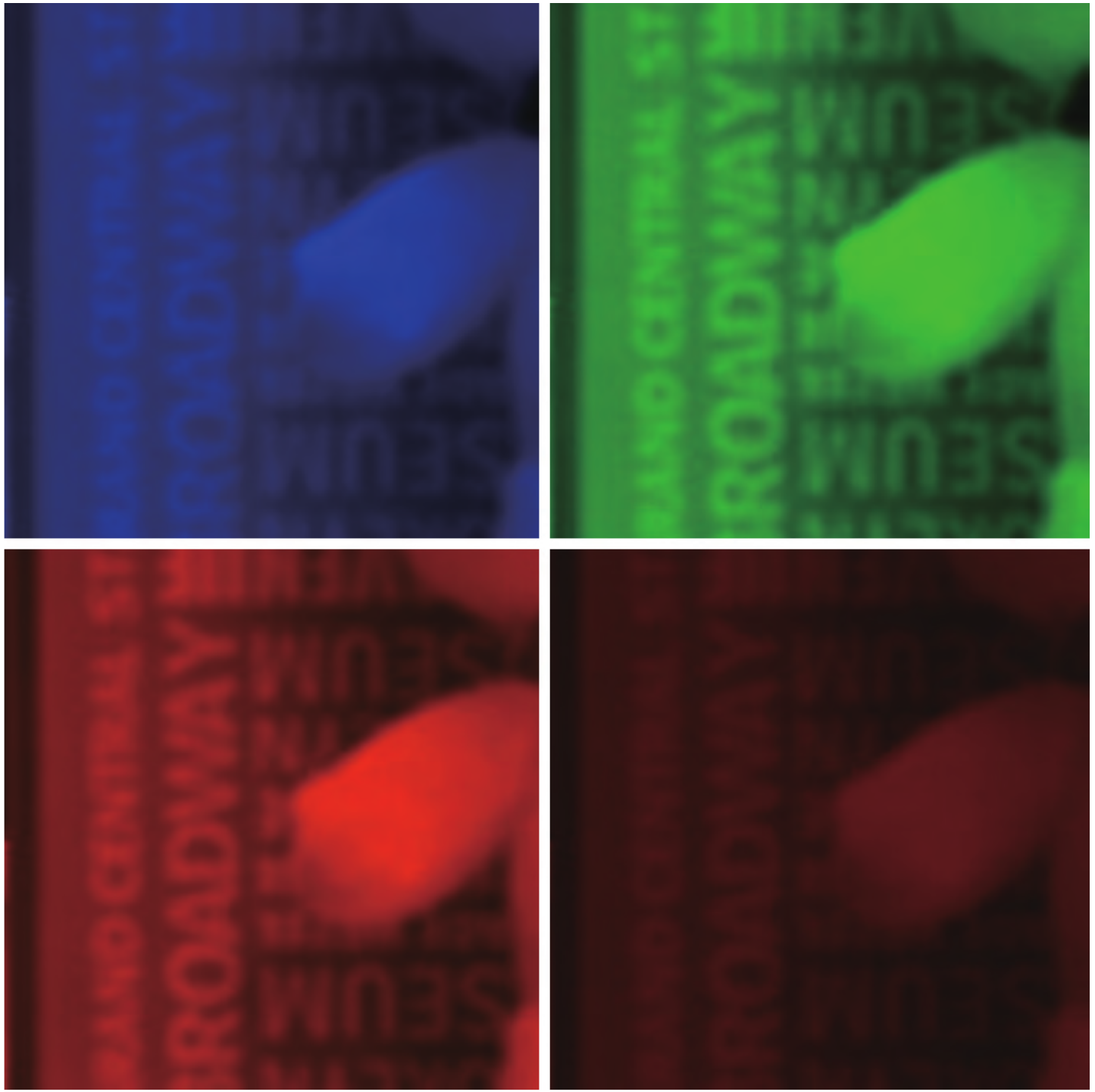}
		
	\end{minipage}
	\begin{minipage}[b]{.25\linewidth}
  \centering
  \centerline{$d_1$}\medskip
	\end{minipage}
\hfill
	\begin{minipage}[b]{0.2\linewidth}
  \centering
  \centerline{$d_4$}\medskip
	\end{minipage}
\hfill
	\begin{minipage}[b]{.20\linewidth}
  \centering
  \centerline{$d_7$}\medskip
\end{minipage}
\hfill
	\begin{minipage}[b]{0.20\linewidth}
  \centering
  \centerline{$d_{10}$}\medskip
	\end{minipage}
\vspace{-3mm}
	\caption{\textbf{The experimental multi-spectral focal stack result on
	real-captured scene} Top 4 rows: selected input at depth $d_1$, $d_4$, $d_7$, $d_{10}$
	with spectral wavelength 430nm, 520nm, 610nm, 700nm. Bottom: the details of
	restored results.}
	\label{fig:multispec}
\vspace{-3mm}
\end{figure}

\paragraph{Qualitative evaluation.} We also show the results of qualitative evaluations. Fig.~\ref{MinionsPic} shows the comparisons on synthetic data between ground truth and our results side by side. To facilitate comparison, we compute the RGB color images from the ground truth and our recovered multispectral slices. It is obvious that the recovered results are very similar to the ground truth. Besides, we also test the method on real captured data. Fig.~\ref{fig:multispec} is an example of our reconstructed multispectral focal stack on the real captured images. The channels of our results at different depths (i.e. $d_1$, $d_4$, $d_7$ and $d_{10}$) are shown in Fig.~\ref{fig:multispec}. Each row represents a selected wavelength, i.e. 430nm, 520nm, 610nm and 700nm, and the close-ups are shown in the bottom of each figure. From the results, we can see that the proposed method performs well on both fine details and smooth areas.


\section{Conclusion and Discussion}
In this paper, we have proposed a chromatic aberration enlarged camera and an LLT-based reconstruction algorithm for acquiring multispectral focal stacks. The proposed method achieves promising performance in terms of both quantitative and qualitative evaluations in our experiments. 

Limited by the complexity, the proposed method cannot work in real time at the current stage. We will try to further simplify and optimize the algorithm in the future.

\bibliographystyle{IEEEbib}
\bibliography{refs}

\end{document}